\documentclass[12pt]{iopart}

\usepackage{bm}
\usepackage{epsf}
\eqnobysec

\newcommand{\lwig}{\mbox{\;\raisebox{.3ex}
    {$<$}$\!\!\!\!\!$\raisebox{-.9ex}{$\sim$}\;}}
\newcommand{\gwig}{\mbox{\;\raisebox{.3ex}
    {$>$}$\!\!\!\!\!$\raisebox{-.9ex}{$\sim$}}\;}
\newcommand{\lambdabar}{{\hbox{$\lambda$\kern-1.ex\raise+0.45ex\hbox{--}}}}  

\begin{document}

\begin{flushright}
{\large \tt DESY 04-147}
\end{flushright}

\title{Gravitational clustering of relic neutrinos and 
implications for their detection}
\author{Andreas Ringwald and Yvonne Y. Y. Wong
}
\address{Deutsches Elektronen-Synchrotron DESY, D-22607 Hamburg,
Germany}

\begin{abstract}
We study the gravitational clustering of 
big bang relic neutrinos onto existing cold dark matter (CDM) and baryonic 
structures within the flat $\Lambda$CDM model, 
using both numerical simulations and a semi-analytical linear technique, 
with the aim of understanding the neutrinos' clustering
properties for direct detection purposes.  In a comparative analysis,
we find that the linear technique systematically underestimates the 
amount of clustering for a wide range of CDM halo and neutrino masses.  
This invalidates earlier claims of the technique's applicability.
We then compute the approximate phase space distribution of relic neutrinos in 
our neighbourhood at Earth, and estimate the large scale neutrino 
density contrasts within the local Greisen--Zatsepin--Kuzmin zone.
With these findings, we discuss the implications of gravitational
neutrino clustering for scattering-based detection methods, ranging 
from flux detection via Cavendish-type torsion balances, to target 
detection using accelerator
beams and cosmic rays.  For emission spectroscopy via
resonant annihilation of extremely energetic cosmic neutrinos on the relic 
neutrino background, we give new estimates for the expected enhancement in
the event rates in the direction of the Virgo cluster. 
\end{abstract}

\ead{\mailto{andreas.ringwald@desy.de}, \mailto{yvonne.wong@desy.de}}

\maketitle

\section{ \label{intro} Introduction}

The standard big bang theory predicts the existence of 
$10^{87}$ neutrinos per 
flavour in the visible universe (e.g., \cite{bib:cosmologytext}).  
This is an enormous abundance unrivalled 
by any other known form of matter, falling second only to the cosmic microwave 
background (CMB) photon.  Yet, unlike the CMB photon which boasts its first 
(serendipitous) detection in the 1960s and which has since been observed 
and its properties 
measured to a high degree of accuracy in a series of airborne/satellite
and ground based experiments, 
the relic neutrino continues to be elusive in the laboratory. 
The chief reason for this is of course the feebleness of the weak 
interaction.  The smallness of the neutrino mass also makes 
momentum-transfer-based detection methods highly impractical.
At present, the only evidence for the relic neutrino comes from 
inferences 
from other cosmological measurements, such as big bang nucleosynthesis 
(BBN) and CMB together with large scale structure (LSS) data (e.g., 
 \cite{bib:hannestadreview}). 
Nevertheless, it is difficult to accept that these neutrinos will never be
detected in a more direct way.

In order to design possible direct, scattering-based detection methods, 
a precise knowledge 
of the phase space distribution of  relic neutrinos is indispensable. 
In this connection, it is important to note that an oscillation interpretation
of the atmospheric 
and solar neutrino  data (e.g., \cite{bib:atmospheric}) 
implies that at least two
of the neutrino mass eigenstates are nonrelativistic today. These neutrinos
are subject to gravitational clustering on existing cold dark 
matter (CDM) and baryonic structures,  
possibly causing the local neutrino number 
density to depart from the standard value of 
$\bar{n}_{\nu} = \bar{n}_{\bar{\nu}} \simeq 56 \ {\rm cm}^{-3}$, 
and the momentum 
distribution to deviate from the relativistic Fermi--Dirac function.

In this paper, we develop a method that will allow us to predict the 
phase space 
distribution of relic neutrinos in our local neighbourhood at Earth
($\sim 8 \ {\rm kpc}$ from the Galactic Centre), 
as well as in outer space. The method systematically takes into account 
gravitational clustering of relic neutrinos
on scales below $\sim 5 \ {\rm Mpc}$, and can be applied to the 
complete range of experimentally and observationally consistent neutrino 
masses. With these predictions, we determine the precise implications
of relic neutrino clustering for future direct search experiments.
To this end, we note that in earlier studies of
relic neutrino direct detection, the neutrino number density in our local
neighbourhood is either assumed to be unrealistically large, or simply left 
as a free parameter (e.g., \cite{Shvartsman:sn,Smith:jj,Duda:2001hd}).  
With the emergence of the concordance flat $\Lambda$CDM
model as the cosmological model of choice, today we are in a position to 
compute the relic neutrino phase space distribution
within a well defined cosmological framework, and to contemplate 
again the prospects for their direct detection in a definitive way.  
Our studies
here will also be useful for  such investigations as relic
neutrino absorption~\cite{Weiler:1982qy,Weiler:1983xx,bib:absorption} 
and emission~\cite{Fargion:1999ft,Weiler:1999sh,Yoshida:1998it,Fodor:2001qy,Fodor:2002hy} spectroscopy.

The standard procedure for any gravitational clustering investigation 
is to solve the $(1+3+3)$-dimensional Vlasov, or collisionless Boltzmann,
equation using $N$-body techniques (e.g., 
\cite{bib:hockney,bib:edfw,bib:bertschingernbody,bib:klypin}).   
However, these techniques are 
computationally very expensive
and necessarily come with 
limited resolutions.  In the context of the cold+hot dark matter (CHDM) 
model, earlier $N$-body studies involving neutrinos probe their kinematic 
effects on structure formation from cluster and galaxy abundances on 
large scales (e.g., \cite{bib:ma&bertschinger,bib:jing&fang,Croft:1994vu,Walter:1995wk}), 
to halo properties on small scales \cite{bib:kofman}. While the CHDM model 
has fallen out of favour in recent years (see, however,  
\cite{bib:sarkar}), it is instructive to note
that the halo simulation of \cite{bib:kofman}  
has a formal resolution of 
only $\sim 100 \ {\rm kpc}$.  This is clearly inadequate for our 
considerations, where the scale of interest is of order $1 \ {\rm kpc}$.

In the context of the flat $\Lambda$CDM model, Singh and Ma (hereafter, SM) 
presented a novel approximate method to probe the accretion
of neutrinos onto CDM halos at scales below $\sim 50 \ {\rm kpc}$
\cite{bib:singh&ma}.  
The salient feature of this study
is their use of parametric 
halo density profiles from high resolution, pure $\Lambda$CDM 
simulations as an external input, while the neutrino component is treated
as a small perturbation whose clustering depends on the CDM halo profile, 
but is too small to affect it in return.  Implementation of this
approximation requires  the neutrino mass density $\rho_{\nu}$ to be 
much smaller than its CDM counterpart $\rho_m$.  
On cosmological scales, we know now from LSS data that the ratio
$\rho_{\nu}/\rho_m = \Omega_{\nu}/\Omega_{m}$ is at most 
$\sim 0.2$ \cite{bib:hannestadreview}.  
On cluster/galactic scales, neutrino free-streaming ensures 
that $\rho_{\nu}/\rho_m$ always remains smaller than its cosmological
counterpart \cite{bib:kofman}. 
Thus the approximation scheme, so far, is sound.  Furthermore, 
in order to track the neutrino density fluctuations in
the most effortless way, SM 
employed the linearised Vlasov equation instead of its full version.
Unfortunately, linear methods are known to break down 
when the density fluctuations reach the order of unity.  
Indeed, in their two trial runs with CHDM parameters, the
linear results of SM compare favourably with 
$N$-body results of
\cite{bib:kofman} in the outer part of the halo,
where the neutrino overdensity is relatively low.
The denser inner parts ($\! \lwig 1 \ {\rm Mpc}$), however, 
show marked disagreement.  This discrepancy renders 
SM's claim 
that their complete prescription is able to probe  neutrino
clustering on sub-galactic scales  doubtful.

In the present investigation, we adopt one of the more attractive 
features of SM's study, namely, the use of parametric 
halo profiles as an external input.
However, we improve upon their analysis by 
solving the Vlasov equation in its (almost) full glory utilising a 
restricted, $N$-$1$-body (pronounced: EN-ONE-BODY) 
method based on the following observation: 
In the limit 
$\rho_{\nu} \ll \rho_m$ and the CDM contribution dominates the 
total gravitational potential, not only will the CDM halo be gravitationally 
blind to the neutrinos, the neutrinos themselves will also have
negligible gravitational 
interaction with each other.  This allows us to track them one 
particle at a time
in $N$ independent simulations, 
instead of following $N$ particles 
simultaneously in one single run, as  in a conventional $N$-body 
study.  An obvious advantage of our $N$-$1$-body technique is that it
requires virtually no computing power when compared with a full scale $N$-body 
simulation with the same, large $N$ ($\! \gwig 10^{6}$). 
It is also less time-consuming since we have 
done away with the need for a gravity solver (the core of
all $N$-body techniques).  In addition, restricted methods such as ours do not 
suffer from spurious two body relaxation, and hence do not 
require the introduction of an artificial softening length that is mandatory
in conventional $N$-body studies.   
Lastly, we note  that restricted methods have been 
used extensively in the studies of galaxy interactions 
(e.g., \cite{bib:sellwood}), and, when properly motivated, 
should not be seen as inferior to full scale
$N$-body techniques.

As a closing remark, let us stress again that our purpose here is not to
investigate the effects of neutrino mass on cosmology, but rather
to address some simple questions such as how many relic 
neutrinos can we realistically expect to find in this very space we occupy,
what kind of energies do they have, where in the universe can we expect to
find the highest concentration of relic neutrinos,
etc., given what we know today about cosmology.
In this regard, the analysis we
present here  is most exhaustive.

The paper is organised as follows.  We begin in section~\ref{cosm} 
with an assessment of the current observational constraints on the relic 
neutrino background.  In section~\ref{vlasov}, 
we introduce the Vlasov equation which is used to track the phase space
distribution of the  neutrinos.  
Section~\ref{profiles} contains a brief discussion of the halo density 
profiles to be employed in our calculations.  
In section~\ref{onebody}, we solve the Vlasov equation for the halo models 
using our improved $N$-$1$-body method for a variety of halo and neutrino
masses.  In section~\ref{linear}, we compute for the same halo models and 
neutrino masses the neutrino overdensities using the linear method of SM 
and examine its validity.  Section~\ref{milkyway} deals exclusively
with relic neutrinos in the Milky Way, and in particular their phase space 
distribution in our immediate vicinity.
We discuss in section~\ref{detection} the implications of our findings 
for scattering-based detection methods, 
and we conclude in section~\ref{conclusion}.

\section{\label{cosm} Observational constraints on the relic 
neutrino background}

Taking as our basis (i) the flat $\Lambda$CDM model with 
$(\Omega_{m,0},\Omega_{\Lambda,0}) \sim (0.3,0.7)$ and Hubble parameter
$h \sim 0.7$,
(ii) neutrino mass splittings inferred from the solar and atmospheric data, 
$(\Delta m^2_{\rm sun}, \Delta m^2_{\rm atm}) 
\sim (10^{-5},10^{-3}) \ {\rm eV}^2$, and (iii) the 
invisible  $Z$ decay width from LEP which constrains the number of 
$SU(2)$ doublet
neutrinos to three \cite{bib:lep},
a minimal theory of neutrino clustering is 
fixed only by the absolute masses of the neutrinos  $m_{\nu}$.  
The current laboratory limit
from tritium $\beta$ decay experiments  is $m_{\nu} < 2.2 \ {\rm eV}$ 
($2 \sigma$) \cite{bib:mainz,bib:troitsk},  
and should  improve to $\sim 0.35 \ {\rm eV}$  
with the upcoming KATRIN experiment \cite{bib:katrin}.
Cosmology also provides a  constraint on $m_{\nu}$. 
For three degenerate species, 
an upper bound of $\sum m_{\nu} < 1.7 \ {\rm eV}$ 
($2 \sigma$) \cite{bib:sdss1,bib:fogli,bib:hannestad2003} 
has been inferred from a combined analysis of the CMB anisotropy from WMAP
\cite{bib:wmapweb} and galaxy clustering from
SDSS (SDSS-gal) \cite{Tegmark:2003uf} (or from 2dFGRS \cite{bib:2dfgrs}), 
together with 
an HST prior on the Hubble parameter \cite{bib:hst}. 
(Reference \cite{bib:fogli} uses also SNIa \cite{bib:sn1a}).\footnote{The mass 
splittings
inferred from the solar and atmospheric neutrino 
oscillation experiments imply that the three mass eigenstates 
are quasi-degenerate when $m_{\nu} \gg \sqrt{\Delta m_{\rm atm}^2}$.} 
Adding to the fit galaxy  bias
\cite{bib:seljak1} and Ly$\alpha$ forest analyses 
can tighten the constraint
to $\sum m_{\nu}< 0.42 \ {\rm eV}$ \cite{bib:seljak2} (see also 
\cite{bib:wmapmass,bib:elgaroy,bib:crottymass,bib:bargermass}), 
although the robustness of 
these additional inputs is still contentious.  Weak lensing of galaxies
\cite{bib:kev} or of the CMB \cite{bib:kaplinghat} will provide
an alternative probe for the cosmological implications of massive neutrinos.

While constraints from cosmology are interesting in their own right,
they are also highly model dependent, and degeneracies abound.
For instance, if tensors and running of the scalar
spectral index are allowed,
the last $\sum m_{\nu}$ bound relaxes to $0.66 \ {\rm eV}$ \cite{bib:seljak2}.
Another possibility is an interplay between
$m_{\nu}$ and $N_{\nu}$, where
$N_{\nu}$ is the {\it effective} number of 
thermalised fermionic degrees of freedom present in the radiation-dominated 
era, such that increasing $N_{\nu}$ actually weakens the bound
on $m_{\nu}$ \cite{bib:hannestad2003}.
For example, a $N_{\nu}=6$ model  
receives a CMB+LSS+priors constraint of 
(i) $\sum m_{\nu} < 2.7 \ {\rm eV}$, if all
six particles are equally massive, (ii) $\sum m_{\nu} < 2.1 \ {\rm eV}$,
if three are massive and the others exactly massless, and (iii)
$\sum m_{\nu} < 4.13 \ {\rm eV}$, if only one is massive 
\cite{bib:hannestad&raffelt}.
Currently,
$1.4 \leq N_{\nu} \leq 6.8$ is allowed by CMB+LSS+priors
\cite{bib:hannestad2003,bib:elgaroy,bib:crotty,bib:pierpaoli,bib:barger2003}. 
Future CMB experiments such as Planck will be sensitive to
$\Delta N_{\nu} \sim 0.2$ \cite{bib:bowen,bib:bashinsky}

Lastly, we note that BBN prefers  
$1.84 \leq N_{\nu} \leq 4.54$ ($2 \sigma$),
in the absence of a $\nu_e$ chemical potential $\zeta_{\nu_e}$ \cite{Cyburt:2004yc}
(see also \cite{bib:barger2003,bib:cuoco}).
Allowing for a  nonzero $\zeta_{\nu_e}$ weakens  the bounds to
$1.3 \leq N_{\nu} \leq 7.1$ 
for $-0.1 \leq \zeta_{\nu_e} \leq 0.3$ \cite{bib:bargerdbbn}.
There is no lack of candidates in the literature for these extra 
$N_{\nu}-3$ degrees of freedom.  We shall not list them here.  
What is certain, however, is that they  cannot take the
form of very large chemical potentials 
in the $\nu_{\mu,\tau}$ sectors, since large neutrino 
mixing inferred from oscillation experiments ensures that  
$\zeta_{\nu_e} \sim \zeta_{\nu_{\mu}} \sim \zeta_{\nu_{\tau}} < 0.3$,
too small to be a significant source of $N_{\nu}$
\cite{bib:dhpprs,bib:yyy,bib:abb}.

In the present analysis, we assume the neutrinos to constitute
exactly three thermalised fermionic degrees of freedom, and 
adopt a conservative mass bound of
\begin{equation}
\label{eq:massbound}
m_{\nu} \lwig 0.6 \ {\rm eV},
\end{equation}
corresponding to the $N_{\nu}=3$ constraint from the WMAP+SDSS galaxy cluster
analysis \cite{bib:sdss1}. 
Alternatively, (\ref{eq:massbound}) may
be interpreted as a restrictive bound for models with extra, non-neutrino
relativistic particles ($N_{\nu}>3$), or with a significant
running spectral index, as discussed earlier.

\section{\label{vlasov} Vlasov equation}

A system consisting of several types of weakly interacting,
self-gravitating particles [e.g., CDM plus neutrinos] 
may be
modelled as a multi-component collisionless gas whose phase space
distributions $f_i(\bm{x},\bm{p},\tau)$ obey the Vlasov equation (e.g.,
\cite{bib:klypin,bib:bertschinger}),
\begin{equation}
\label{eq:vlasov} \frac{D f_i}{D \tau} \equiv \frac{\partial
f_i}{\partial \tau} + \dot{\bm x} \cdot \frac{\partial
f_i}{\partial {\bm x}} + \dot{\bm p} \cdot \frac{\partial
f_i}{\partial {\bm p}} = 0.
\end{equation}
The single-particle phase density $f_i({\bm x}, {\bm p}, \tau)$ is
defined so that $d N_i=f_i \ d^3x \  d^3p$ is the number of $i$
particles in an infinitesimal phase space volume element.  The
variables
\begin{equation}
\label{eq:comoving}
{\bm x} = {\bm r}/a(t), \qquad {\bm p} = a m_i \dot{\bm x}, \qquad d
\tau = dt /a(t),
\end{equation}
are the comoving distance, its associated conjugate momentum, and
the conformal time  respectively, with  $a$ as the scale factor and 
$m_i$ the mass of the $i$th particle species.
All temporal and spatial
derivatives are taken with respect to comoving coordinates,
i.e.,
$\dot{}\equiv \partial/\partial \tau$, $\nabla \equiv
\partial/\partial {\bm x}$.\footnote{Unless otherwise indicated, 
we shall be using comoving 
spatial and temporal quantities throughout the present work.  
Masses and densities, however, are always physical.} 
In the nonrelativistic, Newtonian limit, equation (\ref{eq:vlasov}) is
equivalent to
\begin{equation}
\label{eq:vlasov2} \frac{\partial f_i}{\partial \tau} + \frac{\bm
p}{a m_i} \cdot \frac{\partial f_i}{\partial {\bm x}} - a m_i
\nabla \phi \cdot \frac{\partial f_i}{\partial {\bm p}} = 0,
\end{equation}
with the Poisson equation
\begin{equation}
\nabla^2 \phi = 4 \pi G a^2 \sum_i \overline\rho_i(\tau)
\delta_i ({\bm x}, \tau), \label{eq:poisson}
\end{equation}
\begin{equation}
\delta_i(\bm{x},\tau) \equiv   \frac{\rho_i({\bm x}, \tau)}{
\overline\rho_i(\tau)} -1,  \qquad \rho_i (\bm{x},\tau) =
\frac{m_i}{a^3} \int d^3p \ f_i (\bm{x},\bm{p}, \tau), \label{eq:fluctuations}
\end{equation}
relating the peculiar gravitational potential $\phi (\bm{x},\tau)$
to the density fluctuations $\delta_i (\bm{x},\tau)$ with respect
to the physical mean $\bar{\rho}_i(\tau)$.

The Vlasov equation expresses conservation of phase space density $f_i$ 
along each characteristic $\{{\bm x}(\tau),{\bm p}(\tau)\}$ given by
\begin{equation}
\label{eq:characteristic}
\frac{d \bm{x}}{d \tau} = \frac{\bm{p}}{a m_i}, \qquad \frac{d \bm{p}}{d \tau}
= - a m_i \nabla \phi.
\end{equation}
The complete set of characteristics coming through every point in phase 
space is thus exactly equivalent to equation (\ref{eq:vlasov}).  
It is generally 
not possible to follow the whole set of characteristics, but the evolution
of the system can still be traced, to some extent, if we follow a 
sufficiently large but still manageable sample selected from the 
initial phase space distribution.  This forms the basis of particle-based
solution methods.

\section{\label{profiles} Halo density profiles and other preliminary concerns}

A ``first principles'' approach to neutrino clustering
requires the simultaneous solution of the Vlasov equation (\ref{eq:vlasov}) 
[or, equivalently, the equations for the characteristics 
(\ref{eq:characteristic})]
for both the CDM and the neutrino components. In our treatment, however,
we assume only the CDM component $\rho_m$ contributes to $\phi$ in the 
Poisson equation (\ref{eq:poisson}), and $\rho_m$ to be
completely specified by halo density profiles from high resolution
$\Lambda$CDM simulations.  We provide in this section further
justifications for this approach, as well as a brief discussion on the 
properties of the halo density profiles to be used in our analysis.

It is well known that after they decouple from
the cosmic plasma at $T \sim \ 1 \ {\rm MeV}$, light neutrinos 
($m_\nu \ll 1 \  {\rm keV}$) have too much thermal velocity to cluster on
small scales via gravitational instability in the early stages of structure 
formation.  Accretion onto CDM protoclusters becomes possible only 
after the neutrino velocity has dropped below the velocity 
dispersion of the protoclusters. The mean velocity of the unperturbed 
neutrino distribution has a time dependence of
\begin{equation}
\label{eq:fdspeed}
\langle v \rangle \simeq 1.6 \times 10^2 \ (1+z) \left(\frac{\rm eV}{m_\nu} \right) 
\ {\rm km} \ 
{\rm s}^{-1},
\end{equation}
where $z$ is the redshift.  A typical
galaxy cluster has a velocity dispersion of about 
$1000 \ {\rm km \ s}^{-1}$ today; a typical galaxy, about 
$200 \ {\rm km \ s}^{-1}$.
Thus, for sub-eV neutrinos, clustering on small scales can only have been 
a $z \lwig 2$  event. 

On the other hand, based on a systematic $N$-body study of halo
formation in a 
variety of hierarchical clustering cosmologies, 
Navarro, Frenk and White (hereafter, NFW) argued in 1996 that density
profiles of CDM halos conform to a universal shape, generally 
independently of the halo mass,
the cosmological parameters and the initial conditions 
\cite{Navarro:1995iw,bib:nfw}.  
This so-called NFW profile has a two-parameter functional form  
\begin{equation}
\label{eq:nfw}
\rho_{\rm halo}(r) = \frac{\rho_s}{(r/r_s) (1 + r/r_s)^2},
\end{equation}
where $r_s$ is a characteristic inner radius, and $\rho_s=4 \rho(r_s)$ 
a corresponding inner density.  These parameters $r_s$ and $\rho_s$
are determined by the halo's virial mass $M_{\rm vir}$ and a 
dimensionless concentration
parameter $c$ defined as
\begin{equation}
\label{eq:concentration}
c \equiv \frac{r_{\rm vir}}{r_s},
\end{equation}
where $r_{\rm vir}$ is the virial radius, within which lies $M_{\rm vir}$ 
of matter with an average density equal to $\Delta_{\rm vir}$ times 
the mean matter density  $\bar{\rho}_m$ at that redshift, i.e.,
\begin{eqnarray}
\label{eq:mvir}
M_{\rm vir}  &\equiv &  \frac{4 \pi}{3} \Delta_{\rm vir} 
\bar{\rho}_m a^3 r_{\rm vir}^3 
= \frac{4 \pi}{3} \Delta_{\rm vir} 
\bar{\rho}_{m,0} r_{\rm vir}^3 \nonumber \\
&=& 4 \pi \rho_s a^3 r_s^3 \left[ \ln(1+c)-\frac{c}{1+c} \right],
\end{eqnarray}
where $\bar{\rho}_{m,0}$ is the present day mean matter density.
The factor $\Delta_{\rm vir}$ is usually taken to be the overdensity
predicted by the dissipationless spherical top-hat collapse model 
$\delta_{\rm TH}$, which takes on a value of $\sim 178$ for an 
Einstein--de Sitter cosmology,
while the currently favoured $\Lambda$CDM model has 
$\delta_{\rm TH} \simeq 337$ at 
$z=0$.\footnote{In the original work of NFW 
\cite{Navarro:1995iw,bib:nfw},
$r_{\rm vir}$ is taken to be the radius $r_{200}$, within which the 
average density is $200$ times the {\it critical density} of the universe,
irrespective of the cosmological model in hand.}  

Within this framework, any halo density profile is completely specified 
by its virial mass and concentration via 
equations (\ref{eq:nfw}) to (\ref{eq:mvir}).  Indeed, the NFW profile in
its two-parameter form 
generally gives, for quiet isolated halos,
a fit accurate to $\sim 10 \%$ in the range of radii 
$r= 0.01 \to 1 \ r_{\rm vir}$ \cite{bib:klypin}.
Furthermore, NFW argued for  
a tight correlation between $M_{\rm vir}$ and $c$, such that
the mass distribution within a halo is effectively fixed by the 
halo's virial mass
alone.  Later studies support, to some extent, this conclusion
(e.g., \cite{bib:kravtsov1997,bib:jing2000,bib:bullock2001});
halo concentration correlates with its mass, albeit with a significant 
scatter.  The analysis of  
\cite{bib:bullock2001} of $\sim 5000$ halos in the mass range 
$10^{11} \to 10^{14} M_{\odot}$ 
reveals a trend (at $z=0$) described  by
\begin{equation}
\label{eq:cmvir}
c(z=0) \simeq 9 \left(\frac{M_{\rm vir}}{1.5 \times 10^{13} h^{-1} M_{\odot}} 
\right)^{-0.13},
\end{equation}
with a $1 \sigma$ spread about the median of $\Delta(\log  c) =0.18$ 
for fixed $M_{\rm vir}$.  In addition, for a fixed virial mass, the median
concentration parameter exhibits a redshift dependence of
\begin{equation}
\label{eq:cz}
c(z) \simeq \frac{c(z=0)}{1+z}
\end{equation}
between $z=0$ and $z=4$.

In their analysis, SM interpreted the set of equations (\ref{eq:nfw}) to 
(\ref{eq:cz}) as a complete description of an individual halo's evolution
in time.  While we do not completely disagree with this interpretation,
it should be remembered that the relations (\ref{eq:cmvir}) and 
(\ref{eq:cz}) refer only to the behaviour of the statistical mean for
{\it fixed} virial masses, and do not actually describe how individual 
halos accrete mass over  time.  A better motivation for the 
equations' use
comes from
the observation that the physical densities of the inner regions of
individual isolated halos tend to remain very stable over time 
($z \sim 2 \to 0$) \cite{bib:reed}. This behaviour,
as it turns out, can be more or less reproduced by equations
(\ref{eq:nfw}) to (\ref{eq:cz}), if we interpret $M_{\rm vir}$ as the halo's
virial mass today.  Merging subhalos tend to affect 
the main halo's density profile only in its outer region.
Therefore, provided that neutrino clustering becomes important only 
after $z \sim 2$ and the halos have had no major mergers since then, 
we are justified to either use these equations here, or
simply model the CDM halo as a static object in physical coordinates, 
as a first approximation.

Finally, we note that there exists in the literature a number of other 
halo profiles (e.g., \cite{bib:moore1999,bib:jing&suto}), which, in some cases,
provide better fits to simulations than does 
the NFW profile (\ref{eq:nfw}).  However, these profiles generally
differ from (\ref{eq:nfw}) by less than $10 \%$ \cite{bib:klypin} so we do
not consider them in our study.

\section{\label{onebody} $N$-$1$-body simulations}

Using the NFW halo density profile (\ref{eq:nfw}) as an input, 
we find solutions to the Vlasov equation in the limit 
$\rho_{\nu} \ll \rho_m$ by solving the equations for the 
characteristics (\ref{eq:characteristic}).  
We discuss below the basic set-up. Technical details can be found 
in the Appendix.

\subsection{Basic set-up and assumptions}

We model the CDM distribution as follows.  We assume that throughout space is 
a uniform distribution of CDM.  On top of it, 
sits a spherical NFW halo at the origin.  
In order that the halo overdensity 
merges smoothly into the background density, we extend the NFW profile to 
beyond the virial radius.  Furthermore, for convenience, we treat the NFW 
profile as a perturbation $\bar{\rho}_m (\tau) \delta_m(\bm {x},\tau)$, 
rather than a physical density.  This simplification should make very little
difference to the final results, since the halo density is always much
larger than the background density. The halo's properties and its evolution
in time are contained in the set of equations (\ref{eq:nfw}) to 
(\ref{eq:cz}).
For 
the factor $\Delta_{\rm vir}$, we take a time-independent value
$\Delta_{\rm vir} = 200$,
following \cite{bib:ma&fry} and SM.  We choose this somewhat 
uncommon definition so as to facilitate direct
comparisons between the results of SM and those 
of the present study.  However, since the choice of
$\Delta_{\rm vir}$ affects the profile only through $r_s$ [see
equation (\ref{eq:rs}) in the Appendix], we can see 
immediately that using instead the more 
common $\Delta_{\rm vir} = \delta_{\rm TH}$, where
\begin{equation}
\delta_{\rm TH} \simeq \frac{18 \pi^2 + 82 y - 39 y^2}{\Omega_m(z)}, \quad
y = \Omega_m (z) -1, 
\end{equation}
with $\Omega_m(z) = \Omega_{m,0}/(\Omega_{m,0} + \Omega_{\Lambda,0} a^3)$
\cite{bib:bryan&norman}, will make little  difference to
the outcome.

A more important issue is the role played by other CDM structures 
(i.e., other halos and voids) that should realistically be in the 
surroundings.  In our present scheme, these have all been lumped into 
one uniform background so as to preserve the spherical symmetry of the 
problem.  In reality,  these structures will induce tidal forces
and distort the symmetry.  
However, we expect tidal forces to be important only for
clustering in the outer part of the 
halo where the gravitational potential is low and several halos may 
compete for the same neutrinos.  Clustering in the deep potential
well of the inner part, on 
the other hand, should not be seriously affected.

For the neutrinos, we take the initial distribution to be the
homogeneous and isotropic Fermi--Dirac
distribution with no chemical potential,
\begin{equation}
\label{eq:fermidirac}
f_0 (p) = \frac{1}{1 + \exp(p/T_{\nu,0})},
\end{equation}
where $T_{\nu,0}=1.676 \times 10^{-4} \ {\rm eV}$ is the present day
neutrino temperature.  In principle, the chemical potential need not be 
exactly zero.  In fact, a positive chemical potential $\zeta_{\nu}$ 
should improve, to some extent, the 
clustering of neutrinos (as opposed to anti-neutrinos) by providing more
low velocity specimens that cluster more efficiently than do their high
velocity counterparts.  However, this enhancement is necessarily 
accompanied by a suppression of clustering in the anti-neutrino sector, 
for which a negative $\zeta_{\bar{\nu}} = - \zeta_{\nu}$ tends to deplete
the low velocity states.  Currently, the upper bound on 
$\zeta_{\nu}$ is $0.3$, too small to warrant a detailed investigation 
into a possible ``clustering asymmetry''.

We simulate initial momentum values in the range
$0.01 \leq p/T_{\nu,0} \leq 13$, which accounts for more than $99.9 \%$ of
the distribution (\ref{eq:fermidirac}).  The initial spatial positions of 
the neutrinos range from $r=0$, to as far as it takes
for the fastest particles to land within a distance of 
$10 \ h^{-1} {\rm Mpc}$ from the halo centre at $z=0$.  
We consider a sample of three neutrino masses, 
$m_{\nu} = 0.15,0.3,0.6 \ {\rm eV}$, consistent with
the bound (\ref{eq:massbound}),
and a range of halo viral masses, 
$M_{\rm vir}=10^{12} M_{\odot}  \to 10^{15} M_{\odot}$, corresponding 
to halos of the galaxy to the galaxy cluster variety.
All simulations model a 
flat $\Lambda$CDM cosmology,
with the parameters $(\Omega_{m,0},\Omega_{\Lambda,0},h) = (0.3,0.7,0.7)$

The initial spatial and momentum distribution is divided into small 
chunks that move under the external potential of the CDM halo,
but independently of each other. 
A low resolution run 
is first carried out
for each set of $\{m_{\nu},M_{\rm vir}\}$.  All
chunks that end up at $z=0$ inside a sphere of 
radius $10 \ h^{-1} {\rm Mpc}$ centred on the halo are
traced back to their origin, subdivided into smaller chunks, and 
then re-simulated.  The process is repeated  until the
inner $\sim 10\ h^{-1} {\rm kpc}$ is resolved.  The initial redshift is taken 
to be $z=3$.  This should be sufficient, since clustering is not expected 
to be fully under way until $z \sim 2$.
The final neutrino density distribution is constructed from the set of
discrete particles via a kernel density estimation method 
\cite{bib:reed,bib:merritt} outlined in 
the Appendix, with a maximum smoothing length of $\sim 2 \ h^{-1} {\rm kpc}$ 
in the inner $\sim 50 \ h^{-1} {\rm kpc}$ of the halo.

\begin{figure}
\begin{center}
\epsfxsize=15.7cm
\epsfbox{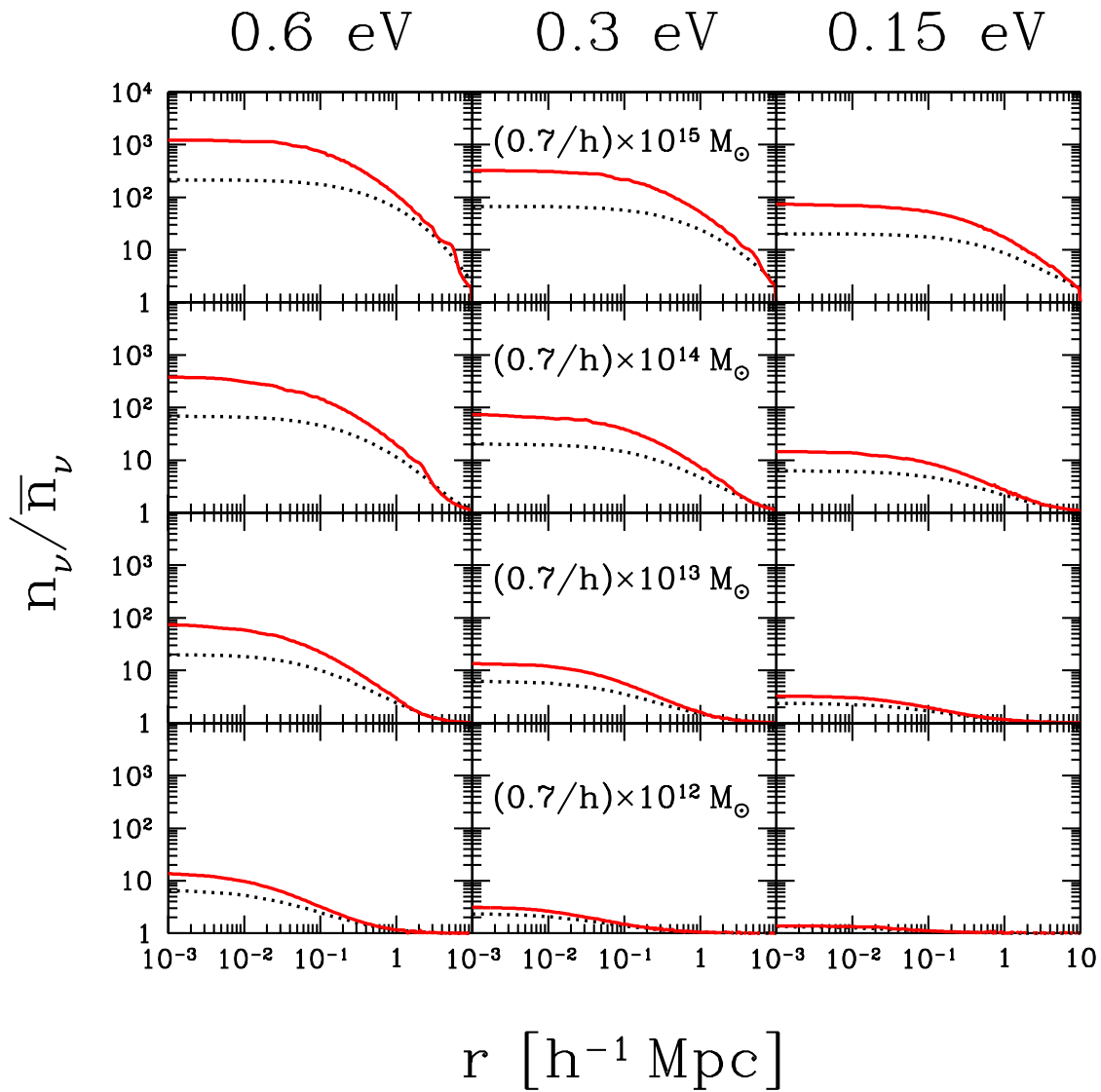}
\end{center}
\caption{\label{fig:overdensities}Relic neutrino number density per 
flavour, $n_{\nu}=n_{\bar{\nu}}$, normalised
to $\bar{n}_{\nu}=\bar{n}_{\bar{\nu}}  \simeq 56 \ {\rm cm}^{-3}$, for 
neutrino masses $m_{\nu}=0.6,0.3,0.15 \ {\rm eV}$ and halo virial masses 
indicated in the figure.  Results from $N$-$1$-body simulations are 
denoted by red (solid) lines.  Dotted lines correspond to overdensities
calculated with the linear approximation.}
\end{figure}

\begin{figure}
\begin{center}
\epsfxsize=15.7cm
\epsfbox{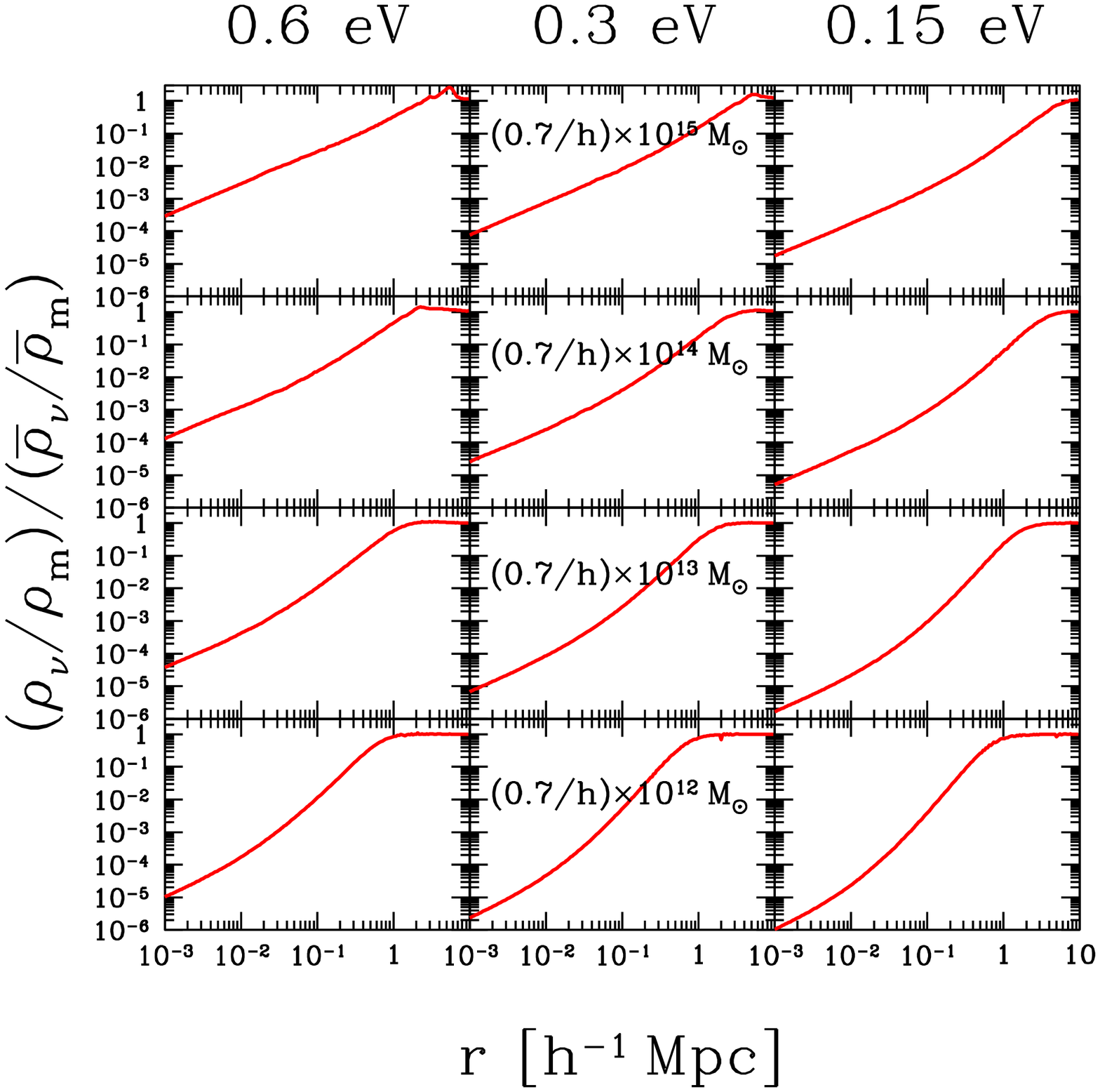}
\end{center}
\caption{\label{fig:nuvscdm}Mass density ratio $\rho_{\nu}/\rho_{m}$ 
normalised to the background mean $\bar{\rho}_{\nu}/\bar{\rho}_m$
obtained from $N$-$1$-body simulations for
various neutrino and halo masses indicated in the figure.}
\end{figure}

\subsection{Results and discussions}

The basic results of our $N$-$1$-body simulations are presented 
in Figure \ref{fig:overdensities}, which shows the neutrino 
overdensities $n_{\nu}/\bar{n}_{\nu}$ for various sets of $\{m_{\nu},
M_{\rm vir} \}$.  A companion figure, Figure 
\ref{fig:nuvscdm},  shows the same results expressed in terms of
the mass density ratio $\rho_{\nu}/\rho_{m}$ normalised to the 
background mean $\bar{\rho}_{\nu}/\bar{\rho}_m$.

The essential features of the curves in Figures 
\ref{fig:overdensities} and \ref{fig:nuvscdm} can be understood
in terms of neutrino free-streaming, which causes the 
$n_{\nu}/\bar{n}_{\nu}$ curves to flatten out at small radii, and 
the mass density ratio $\rho_{\nu}/\rho_{m}$ to drop substantially 
below the background mean.  (The latter feature also provides 
a justification for our $N$-$1$-body method.)
Both $n_{\nu}/\bar{n}_{\nu}$ and $\rho_{\nu}/\rho_m$  
approach their respective
cosmic mean of $1$ and $\bar{\rho}_{\nu}/\bar{\rho}_m$
at large radii.  Similar behaviours have also been
observed in the CHDM simulations of reference \cite{bib:kofman}.

\begin{figure}
\begin{center}
\epsfxsize=15.7cm
\epsfbox{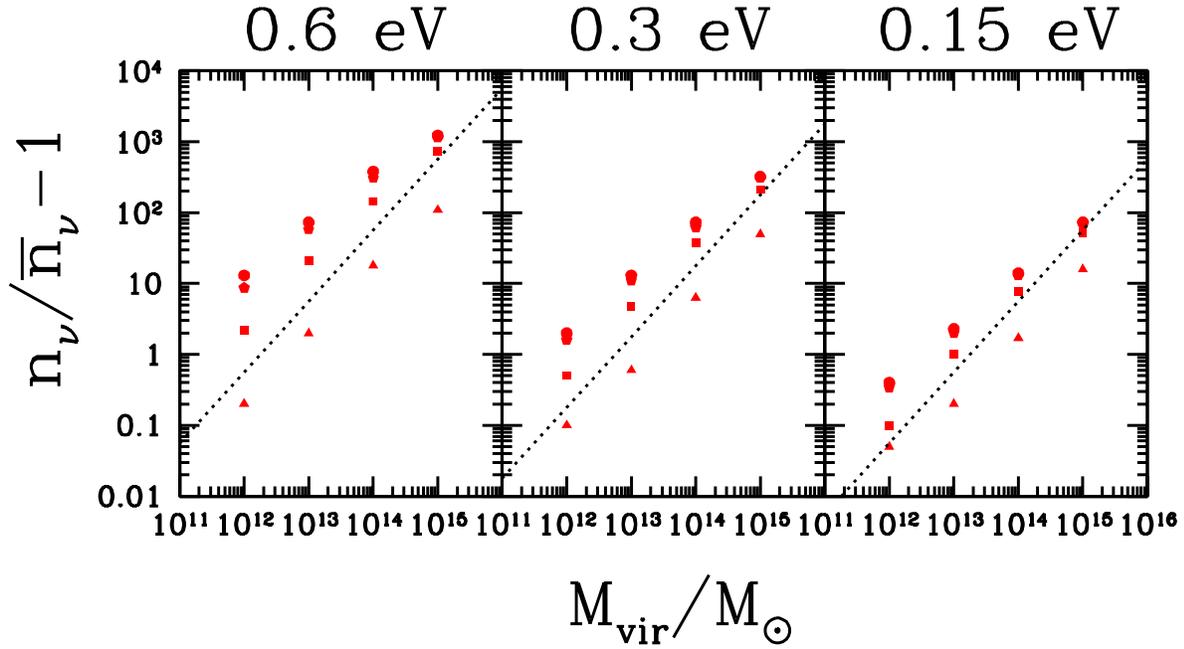}
\end{center}
\caption{\label{fig:halomass}Dependence of the neutrino overdensity on
the halo virial mass for neutrino masses $m_{\nu}=0.6,0.3,0.15 \ {\rm eV}$.  
The circles represent  overdensities at
$1 \ h^{-1}{\rm kpc}$, pentagons at $10 \ h^{-1}{\rm kpc}$, squares at 
$100 \ h^{-1}{\rm kpc}$ and triangles at $1000 \ h^{-1}{\rm kpc}$.
The straight lines ($\propto M_{\rm vir}$) are provided to guide the eye and 
are {\it not} meant to be best fits.}
\end{figure}

\begin{figure}
\begin{center}
\epsfxsize=15.0cm
\epsfbox{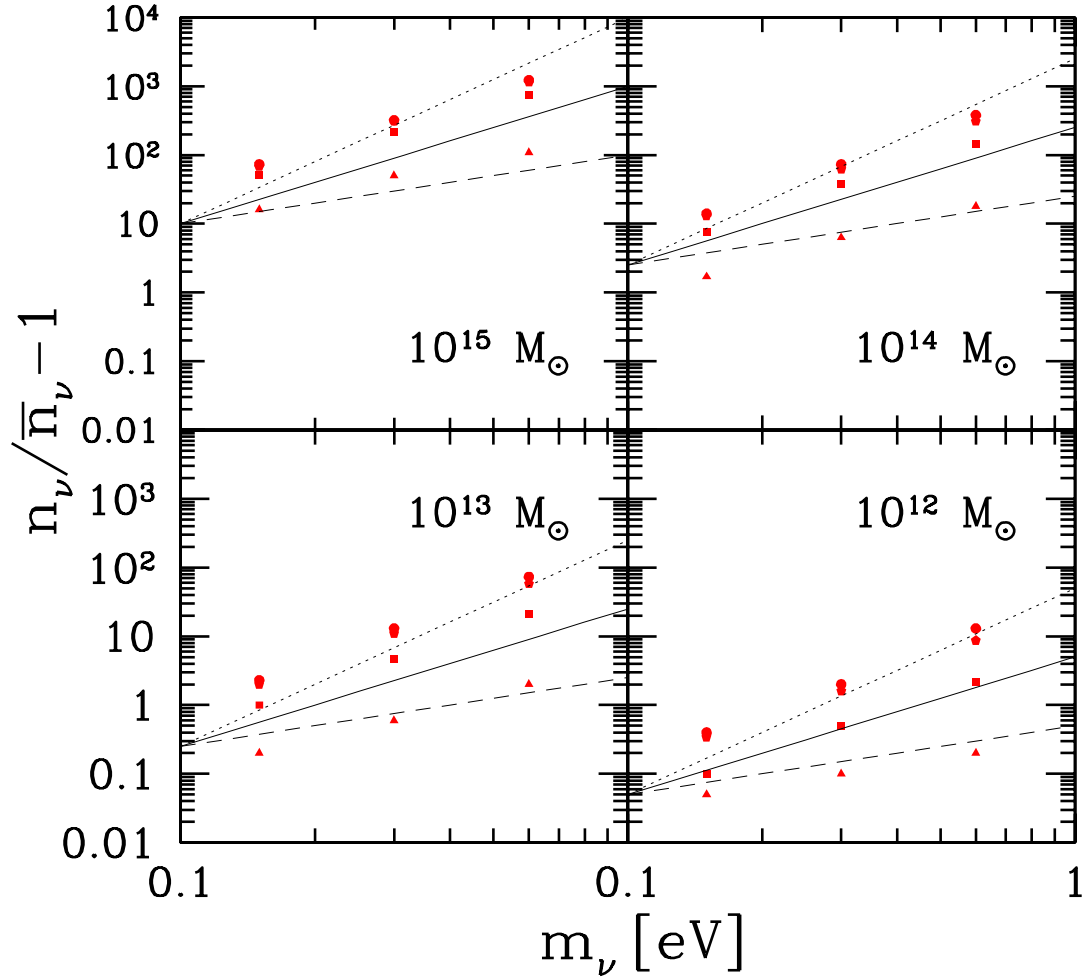}
\end{center}
\caption{\label{fig:msquare}Dependence of the neutrino overdensity on the 
neutrino mass for various halo masses indicated on the plots.  
The circles represent  overdensities at
$1 \ h^{-1} {\rm kpc}$, pentagons at $10 \ h^{-1} {\rm kpc}$, squares at 
$100 \ h^{-1}{\rm kpc}$ and triangles at $1000 \ h^{-1}{\rm kpc}$.
The solid lines correspond to an $m_{\nu}^2$ dependence, dashed lines
an $m_{\nu}$ dependence, and dotted lines an $m_{\nu}^3$ dependence.
These lines are provided to guide the eye and
are {\it not} meant to be best fits.}
\end{figure}

Naturally, clustering improves with increasing neutrino and/or
halo masses.  
In Figure \ref{fig:halomass}, we plot the neutrino
overdensities at $1,10,100,1000 \ h^{-1} {\rm kpc}$ for various neutrino
masses as a function of the halo virial mass $M_{\rm vir}$.  
A similar plot is constructed in Figure \ref{fig:msquare}, 
with the neutrino mass $m_{\nu}$ as the independent variable.
In the former, the quantity $n_{\nu}/\bar{n}_{\nu}-1$ is seen to be 
roughly proportional to $M_{\rm vir}$ for a fixed radius and 
a fixed neutrino
mass.  
The $n_{\nu}/\bar{n}_{\nu}-1$ versus $m_{\nu}$ trend
in Figure \ref{fig:msquare} is more difficult to quantify.
A roughly $m_{\nu}^2$ dependence can be discerned for some
fixed halo masses at some fixed radii.  Other combinations, 
however, display noticeably different 
behaviours.   Whatever these dependences are, it is 
interesting to note that they are never shallower 
than $\propto m_{\nu}$, or steeper than $\propto m_{\nu}^3$.
As we shall see later, an $m_{\nu}^2$ and an $m_{\nu}^3$ dependence for
the overdensities can both be motivated from theory. The former
can be derived from the linearised Vlasov equation (cf. section~\ref{linear}),
while the latter follows naturally from phase space considerations  
(or the so-called Tremaine--Gunn bound, cf. section~\ref{milkyway}).

Lastly, in order to test the robustness of our results, we (i) push the 
initial redshift of the simulations back to $z=5$, (ii) vary the 
cosmological parameters within their allowed ranges, and (iii) alter the 
time dependences of some of the halo parameters.  In all cases, we find,
as expected, the heavier masses $\{m_{\nu},M_{\rm vir}\}$ 
to suffer more from these variations.  For our heaviest set, 
$\{m_{\nu}=0.6 \ {\rm eV},M_{\rm vir} = 0.7 \times 10^{15} \
h^{-1}M_{\odot}\}$, the neutrino overdensity changes by about 
$10$ to $20 \%$ at small radii, and some $50$ to $100 \%$ at
$r \gwig 5 \ h^{-1} {\rm Mpc}$.  The gap narrows with 
smaller neutrino and halo masses. For galaxy size halos 
($M_{\rm vir} \sim 10^{12} M_{\odot}$), the density
variations with respect to (i), (ii) and (iii) 
are no more than $\sim 10 \%$ everywhere.  Thus
our simulation results are generally quite robust.

\section{\label{linear} Linear approximation}

The linear approximation is often used in the literature to find
approximate solutions to the Vlasov equation.  In the nonrelativistic
limit, the pioneering work of Gilbert \cite{bib:gilbert} has, over the years,
been applied to the study of the pure hot dark matter (HDM) model
\cite{bib:bond}, 
as well as in the analysis of HDM accretion onto nonadiabatic seeds such 
as cosmic strings \cite{bib:turok,bib:watts,bib:stebbins} and onto CDM
halos \cite{bib:singh&ma}. 
The procedure 
consists of first switching to a new time variable
$s \equiv \int a^{-1} d \tau = \int a^{-2} d t$,
and then Fourier transforming the $\bm{x}$-dependent functions,
\begin{equation}
\hat{f} (\bm{k},\bm{p},s)  \equiv \mathcal{F}[f
(\bm{x},\bm{p},s)], \qquad \hat{\phi} (\bm{k},s)   \equiv 
\mathcal{F}[\phi (\bm{x},s)],
\end{equation}
to obtain a new differential equation in Fourier space,
\begin{equation}
\label{eq:vlasovfourier}
\frac{\partial \hat{f}}{\partial s} + \frac{i \bm{k} \cdot
\bm{p}}{m_{\nu}} \hat{f}  - i m_{\nu} a^2  (\bm{k} \ \hat{\phi}  \star
\nabla_{\bm p} \hat{f})=0,
\end{equation}
where 
$\bm{k} \ \hat{\phi}  \star \nabla_p \hat{f} \equiv \int d^3
\bm{k}' \ \bm{k}'\  \hat{\phi} (\bm{k}')  \cdot \nabla_{\bm p}
\hat{f} (\bm{k}-\bm{k}')$
is the convolution product.  The equation is said to be linearised when
one makes the replacement
\begin{equation}
\label{eq:replacement}
\nabla_{\bm{p}} \hat{f}(\bm{k}) \to \nabla_{\bm{p}}f_0(p) \delta(\bm{k}),
\end{equation}
where $\delta(\bm{k})$ is the Dirac delta function, so that the
convolution product becomes a simple scalar product
$\bm{k} \hat{\phi} \cdot \nabla_{\bm{p}} f_0$.  Replacement
(\ref{eq:replacement})  is
valid as long as the condition
$\left| \nabla_{\bm p} (f-f_0) \right| \ll \left| \nabla_{\bm p}
f_0 \right|$ holds.  In practice, however, the quantity $\nabla_{\bm p} f$
is somewhat cumbersome to compute, so the ``rule of thumb'' regarding the
linear approximation is to abandon it as soon as the 
spatial density fluctuation $\delta_{\nu}(\bm{x},s)$, defined in
(\ref{eq:fluctuations}),  exceeds order unity,
as  emphasised in \cite{bib:bertschinger,bib:watts}.  
Nonetheless, the linear theory
has been time and again used beyond this putative limit.  We shall
also apply it to our case, to gain physical insight as well as to
see how it compares with $N$-$1$-body simulations.

Equation (\ref{eq:vlasovfourier}) together with the replacement 
(\ref{eq:replacement}) has a very simple solution,
\begin{eqnarray}
\label{eq:fsolution}
 \hat{f}(\bm{k}, \bm{p},s)& = &
\hat{f}(\bm{k},\bm{p},s_i) e^{-i\bm{k} \cdot \bm{u} (s-s_i)}
 \nonumber \\ && + i m_{\nu}\ \bm{k} \cdot
\nabla_{\bm p} f_0 \int^s_{s_i} ds' a^2(s)
\hat{\phi}(\bm{k},s') e^{-i \bm{k} \cdot \bm{u} (s-s')},
\end{eqnarray}
where $\bm{u} = \bm{p} /m_{\nu}$, $s_i$ is some initial time, and
we take the initial phase space distribution to be isotropic and 
homogeneous in space, i.e.,
$\hat{f}(\bm{k},\bm{p},s_i) = \delta (\bm{k}) f_0(p)$. 
The neutrino number density per Fourier mode relative to the mean
density is obtained by integrating (\ref{eq:fsolution}) over
momenta $\bm{p}$,
\begin{eqnarray}
\label{eq:nudensity} \frac{\hat{n}_{\nu} (\bm{k},s)}{\bar{n}_{\nu}(s)} &
= & \frac{a^{-3} \int d^3p \ \hat{f}(\bm{k},\bm{p},s)}{a^{-3} \int
d^3 p \ f_0(p)} \equiv \frac{1}{\bar{n}_{\nu,0}} \int d^3 p \ \hat{f}
(\bm{k},\bm{p}, s) \nonumber
\\  & = &  \delta(\bm{k})  - k^2 \int^{s}_{s_i} ds'
a^2(s') \hat{\phi}(\bm{k},s') (s-s') F \left[\frac{k
(s-s')}{m_{\nu}} \right],
\end{eqnarray}
with
\begin{equation}
\label{eq:fq}
F(q) \equiv \frac{1}{\bar{n}_{\nu,0}}  \int d^3p \ e^{-i \bm{p} \cdot \bm{q}}
f_0 (p).
\end{equation}
The correct form for $f_0(p)$ should be the relativistic Fermi--Dirac 
function (\ref{eq:fermidirac}), which gives for $F(q)$ a series 
representation \cite{bib:watts}
\begin{equation}
F(q) = \frac{4}{3 \zeta(3)} \sum^{\infty}_{n=1} (-1)^{n+1}
\frac{n}{(n^2 + q^2 T_{\nu,0}^2)^2},
\end{equation}
where $\zeta(3) \simeq 1.202$ is the Riemann zeta function.
However, in order to simplify calculations and/or to gain 
physical insight, other forms of $f_0(p)$ have also appeared in the
literature.  Some are listed in Table \ref{table:F(q)}, along
with their corresponding $F(q)$. 

The solution to the 
Poisson equation (\ref{eq:poisson}) in Fourier space is
\begin{equation}
\hat{\phi}(\bm{k},s) = - \frac{4 \pi G a^2 \bar{\rho}_m (s) 
\hat{\delta}_m (\bm{k},s)}{k^2} =  
- \frac{4 \pi G  \bar{\rho}_{m,0}  
\hat{\delta}_m (\bm{k},s)}{a k^2}.
\end{equation}
Substituting this into  (\ref{eq:nudensity}) and using the definition
$\hat{\delta}_{\nu} (\bm{k},s) \equiv 
\hat{n}_{\nu}(\bm{k},s)/\bar{n}_{\nu}(s) -\delta(\bm{k})$, we obtain
for the neutrino density fluctuations
\begin{equation}
\label{eq:perturbation}
\hat{\delta}_{\nu} (\bm{k},s) \simeq 4 \pi G \bar{\rho}_{m,0}
\int^{s}_{s_i} ds' a(s') \hat{\delta}_m(\bm{k},s') (s-s') F
\left[\frac{k (s-s')}{m_{\nu}} \right].
\end{equation}
This is the ``master equation'' for the linear approach.
We solve equation ({\ref{eq:perturbation}) numerically for a variety of 
neutrino and halo masses.  The results are presented in Figure 
\ref{fig:overdensities}, alongside their $N$-$1$-body counterparts.

\begin{table}
\caption{\label{table:F(q)} Some distribution functions $f_0(p)$ and 
their corresponding $F(q)$ [equation (\ref{eq:fq})] that have appeared
in the literature.  The series solution for $F(q)$ for the 
relativistic Fermi--Dirac (FD) function was first derived in 
\cite{bib:watts}.  A Maxwell--Boltzmann (MB) type 
distribution was adopted
in \cite{bib:turok}.  The last family of distributions, characterised
by $F(q)$'s exponential form, 
appears in \cite{bib:bertschinger} and \cite{bib:setayeshgar}.}
\begin{indented}
\item[]\begin{tabular}{@{}lll}
\br Distribution & $f_0(p)$ & $F(q)$ \\ \mr Relativistic FD & 
$[1+\exp(p/T_{\nu,0})]^{-1}$ & $\frac{4}{3 \zeta(3)} \sum^{\infty}_{n=1}
(-1)^{n+1} n(n^2 + q^2 T_{\nu,0}^2)^{-2}$ \\ MB & 
$\exp(-p/T_{\nu,0}) $  & $ (1 + q^2 T_{\nu,0}^2)^{-2}$ \\ 
$\gamma$ distribution  & $\frac{\bar{n}_{\nu,0}}{\pi^2 (\gamma T_{\nu,0})^3} 
(1+ p^2/\gamma^2 T_{\nu,0}^2)^{-2}$ &
$\exp(-\gamma q T_{\nu,0})$ \\ \br
\end{tabular}
\end{indented}
\end{table}

\subsection{Further approximations and analytical insights}

Before comparing the two approaches,
let us first study the linear approximation for its own sake.
Consider the master equation (\ref{eq:perturbation}).  
In the limit 
$F(q)$ grows much faster than $a(s)$ and $\hat{\delta}_m (\bm{k},s)$,
i.e.,
\begin{equation}
\label{eq:asymptoticcondition}
\frac{k T_{\nu,0}}{m_{\nu}} \gg \frac{1}{a} \frac{d a}{d s} + 
\frac{1}{\hat{\delta}_m} \frac{d \hat{\delta}_m}{d s},
\end{equation}
equation (\ref{eq:perturbation}) may be solved by asymptotic expansion.
The resulting approximate solution looks somewhat messy at
first sight,
\begin{eqnarray}
\hat{\delta}_{\nu} (\bm{k},s) & \simeq &  4 \pi G \bar{\rho}_{m,0}
\left( \frac{m_{\nu}}{k T_{\nu,0}} \right)^2 \frac{2}{3 \zeta (3)} [
\ln(2) a(s) \hat{\delta}_{m} (\bm{k},s) -  \nonumber \\ && 
 a(s_i) \hat{\delta}_m (\bm{k},s_i) \sum^{\infty}_{n=1 } 
(-1)^{n+1} \frac{n}{n^2 + \frac{k^2 T_{\nu,0}^2 }{m_{\nu}^2} (s-s_i)^2} 
], 
\end{eqnarray}
but may be rendered into a physically transparent form if we first cross out  
the second term, which is well justified since the initial  
$a$ and $\hat{\delta}_m$ should always be much  
smaller than the final ones, and then rewrite the expression as   
\begin{equation} 
\label{eq:asymptoticsolution} 
\hat{\delta}_{\nu} (\bm{k},s) \simeq \frac{k_{\rm fs}^2(s)}{k^2} 
\hat{\delta}_{m}(\bm{k},s). 
\end{equation} 
Here, $k_{\rm fs}$ is but the free-streaming wave vector, defined as 
\begin{eqnarray} 
k_{\rm fs} (s) & \equiv & \sqrt{\frac{4 \pi G a(s) 
\bar{\rho}_{m,0}}{c_{\nu,0}^2}} = \sqrt{\frac{4 \pi G a^2(s) 
\bar{\rho}_m(s)}{c_{\nu}^2(s)}} \nonumber \\ 
&\simeq &  1.5 \ \sqrt{a(s) \Omega_{m,0}}\  \left( 
\frac{m_{\nu}}{\rm eV} \right) \ h \ {\rm Mpc}^{-1}, 
\end{eqnarray} 
and we identify 
\begin{eqnarray} 
c_{\nu}(s) & \equiv  & \frac{T_{\nu,0}}{m_{\nu}a (s)} \sqrt{\frac{3 
\zeta(3)}{2 \ln(2)}} \equiv \frac{ c_{\nu,0}}{a(s)} \nonumber \\ 
& \simeq &   \frac{81}{a(s)}\left(\frac{\rm eV}{m_{\nu}}\right) \ {\rm km\ 
s}^{-1} 
\end{eqnarray} 
as the neutrino's characteristic thermal speed. 
 
The functional form of equation (\ref{eq:asymptoticsolution}) already 
tells us something very interesting; large Fourier modes in the neutrino 
density fluctuations are suppressed by a factor proportional to $k^{-2}$  
relative to their CDM counterparts.  This is clearly a manifestation 
of free-streaming, which is responsible for inhibiting the growth of
structures on  scales below 
$\lambda_{\rm fs} \equiv 2 \pi/k_{\rm fs}$.  
Furthermore, $\hat{\delta}_{\nu}$ has  
an $m_{\nu}^2$ dependence through $k_{\rm fs}$, meaning that, 
at small scales,  
a neutrino twice as heavy as another is able to cluster four times  
more efficiently.  This $m_{\nu}^2$ dependence is reflected, approximately, by 
both our linear and $N$-$1$-body results in Figure \ref{fig:overdensities}, 
and is particularly pronounced at small radii.

That equation (\ref{eq:asymptoticsolution}) is a solution of  
(\ref{eq:perturbation}) is contingent upon the satisfaction of the condition  
(\ref{eq:asymptoticcondition}), which requires, for a fixed neutrino
mass, $k$ to be larger than some nominal $k_{\rm min}$ determined by the 
rates  
of change of the scale factor $a$ and of the CDM perturbations  
$\hat{\delta}_m(\bm{k},s)$. 
The rate of change of $a$ is a simple and well defined function of the  
cosmological model.   
The growth rate of $\hat{\delta}_m (\bm{k},s)$, on the other hand,  
is usually more complicated.  However, because our halos are practically  
static in physical coordinates, this rate (in comoving Fourier space)  
can only be at most of the order of the universal expansion rate  
(which comes in through the conversion factor $a$ when we switch the halo
profile from physical to comoving coordinates).  Thus the condition  
(\ref{eq:asymptoticcondition}) is roughly equivalent to 
\begin{equation} 
\label{eq:asymptoticcondition2} 
k \gg k_{\rm min}(s) \sim \frac{m_{\nu}}{T_{\nu,0}} a^2 H(s) \simeq 
2 \sqrt{a \Omega_{m,0} +a^4 \Omega_{\Lambda,0}}  
\left(\frac{m_{\nu}}{\rm eV} \right) h \ {\rm Mpc}^{-1}, 
\end{equation} 
where $H(s)$ is the Hubble expansion parameter at time $s$.  Since
 $k_{\rm min} \sim k_{\rm fs}$ at most times, we see that
equation (\ref{eq:asymptoticsolution}) is indeed applicable to
all $k$ modes larger than the free-streaming wave vector $k_{\rm fs}$.

Unfortunately, the opposite $k \ll k_{\rm fs}$ limit has no simple  
approximate solution because of the complicated  dependence of the 
scale factor $a$ on the new time variable $s$. 
However, we find the following formula to give a decent fit 
to the solution of (\ref{eq:perturbation}) for a wide range of $k$, 
\begin{equation} 
\label{eq:fitting}
\hat{\delta}_{\nu}(\bm{k},s) \simeq 
\frac{k_{\rm fs}^2}{(k_{\rm fs}/\Gamma + k)^2}
\hat{\delta}_m (\bm{k},s) \equiv \hat{K}(k_{\rm fs}^{-1} \bm{k},s) 
\hat{\delta}_m (\bm{k},s),
\end{equation} 
with 
\begin{equation}
\Gamma^2 \equiv \left. \frac{4 \pi G \bar{\rho}_{m,0}}
{\hat{\delta}_m(\bm{k},s)} 
\int^{s}_{s_i} ds' a(s') \hat{\delta}_m(\bm{k},s') (s-s') 
\right|_{\bm{k} \to 0}.
\end{equation}
Typically, $\Gamma \sim 1$, such that for $k \ll k_{\rm fs}$, 
the growth of $\hat{\delta}_{\nu}$ approximately matches that of 
$\hat{\delta}_m$.  Therefore, equation (\ref{eq:fitting}) is roughly 
equivalent to
\begin{equation}
\label{eq:nufilter}
\rho_{\nu}(\bm{x}) \sim k_{\rm fs}^3 K(k_{\rm fs} \bm{x}) 
\star \rho_m(\bm{x}),
\end{equation}
in real space, with $K(\bm{x}) \equiv {\cal F}^{-1}[\hat{K}(\bm{k})]$ 
acting like a normalised filter function with window width 
$k_{\rm fs}^{-1}$, that gently smears out the neutrino density contrasts 
on scales below $\sim k_{\rm fs}^{-1}$ relative to their CDM counterparts.  
We shall be using equation (\ref{eq:nufilter}) again in section~\ref{target}.

\subsection{Comparison with $N$-$1$-body results: when and 
how the linear theory fails} 
 
Comparing the linear results from this section 
with $N$-$1$-body simulations from
section~\ref{onebody}, it is immediately 
clear in Figure~\ref{fig:overdensities} that the former 
systematically underestimates the neutrino overdensities over
the whole range of neutrino and halo masses considered in this study. 
  The discrepancy is most prominent in the 
dense, inner regions ($r \lwig  1 \ h^{-1} {\rm Mpc}$), 
and worsens as we increase
(i) the neutrino mass $m_{\nu}$, and (ii) the halo mass $M_{\rm vir}$.  
The worst case corresponds to 
when both $m_{\nu}$ and $M_{\rm vir}$ are large; the case
of $\{m_{\nu}=0.6 \ {\rm eV}, M_{\rm vir} = 0.7 \times 10^{15} \
h^{-1} M_{\odot} \}$, for example, sees the $N$-$1$-body and the linear
overdensities differ by a factor of about six.
For smaller neutrino and halo masses, concordance between the two
approaches improves as we move to larger radii.  Indeed, for 
$\{m_{\nu}=0.15 \ {\rm eV}, M_{\rm vir} = 0.7 \times 10^{12} \
h^{-1} M_{\odot} \}$, complete agreement is seen throughout the region
of interest.
Upon closer inspection, one finds that the linear theory ceases to be 
a faithful approximation once the neutrino 
overdensity reaches a value of about three or four.  This is of course
fully consistent with the standard lore that perturbative methods fail once
the perturbations exceed unity and nonlinear effects set in.

Can the linear approximation be salvaged?   Yes, provided we impose a 
great deal of smoothing.  In the case of 
$\{m_{\nu}=0.6 \ {\rm eV}, M_{\rm vir} = 0.7 \times 10^{15} \
h^{-1} M_{\odot} \}$, for example, the overdensities computed
from the two different approaches can be reconciled with each other
if we smooth them both over a scale of roughly 
$5 \ h^{-1}{\rm Mpc}$.  Such a large smoothing length
will render the linear method completely useless for the study of
neutrino clustering on sub-galactic scales
(unless of course the neutrino mass is so small that the
overdensity does not exceed unity by much anyway).  But the method can
still be useful for obtaining quick estimates of $n_{\nu}/\bar{n}_{\nu}$
on larger scales for absorption and emission spectroscopy  
calculations (cf. section~\ref{target}).

Finally, we note that the neutrino overdensities in Figure 2 of SM are at odds with
our linear results in Figure
\ref{fig:overdensities}.
This discrepancy cannot be ameliorated by simply supposing that SM have normalised
their neutrino densities for {\it three} flavours to the {\it one} flavour average
$\bar{n}_{\nu} = \bar{n}_{\bar{\nu}} \simeq 56 \ {\rm cm}^{-3}$, since this normalisation
will render some of their results---specifically, 
where $n_{\nu}/\bar{n}_{\nu} <3$---unphysical.
Only a normalisation to {\it three} flavours can give these results a physical meaning,
but at the expense of incompatibility with our Figure \ref{fig:overdensities}, 
as well as with SM's own Figure 3.
We therefore conclude that SM's results as presented in their Figure 2
are erroneous.

\section{\label{milkyway} Relic neutrinos in the Milky Way}

In this section, we consider relic neutrino clustering in the 
Milky Way.  We compute explicitly the number of neutrinos and their 
distribution in momentum space in our local neighbourhood at Earth
($r_{\oplus} 
\sim 8 \ {\rm kpc}$ from the Galactic Centre).  This information is
essential for any direct search experiment.

\subsection{Background, basic set-up and assumptions}

We perform this calculation using the $N$-$1$-body method of
section~\ref{onebody}, 
but with a few modifications to the external potential.  Firstly, we note 
that the central region of the Milky Way ($\! \lwig 10 \ {\rm kpc}$) 
is dominated 
by baryonic matter in the form of a disk, a bulge, and possibly a rapidly
rotating bar \cite{bib:binney&tremaine}.  
Each of these components has its own distinct density profile.
Furthermore, in the standard theory of hierarchical galaxy formation, 
baryons and dark matter are initially well mixed, and collapse together 
to form halos via
gravitational instability.  Galactic structures arise 
when the baryons cool and fall out of the
original halo towards the centre \cite{bib:white&rees}.  
As the baryons condense, their 
gravitational forces tend to pull the dark matter inward, thereby distorting
the inner $\sim 10 \ {\rm kpc}$ of the original halo profile 
(e.g., \cite{bib:blumenthal,bib:flores,bib:mmw,bib:gnedin}).  This kind
of modification to the mass distribution is important for us at 
$r_{\oplus}$.  Fortunately for our calculation,
gravitating neutrinos do not distinguish between halo and baryons.
Therefore, it suffices to use simply the total mass distribution inferred
from observational data (e.g., rotation curves, satellite kinematics, 
etc.), without any detailed modelling of the individual components.

What is still missing, however, is the redshift dependence of the mass 
distribution.  
Unfortunately, we have not been able to find in the literature any simple 
parametric
form for this dependence.
However, mass modelling of the Milky Way \cite{bib:dehnen,bib:klypin2002} 
suggests that
certain observationally acceptable bulge+disk+halo models 
are indeed consistent with the aforementioned theory of baryonic
compression, and
can be traced back to halos {\it originally} of the NFW form by
supposing that the system has undergone a phase of adiabatic 
contraction \cite{bib:blumenthal}.  Thus, for
our investigation, it is probably fair
to think of the NFW profile as the initial mass distribution, and the
evolution as a smooth transition from this initial distribution 
to the present day one.

Instead of modelling this transition, however, our strategy here 
is to conduct two series of simulations, one for the 
present day mass distribution of the Milky Way (MWnow) 
which we assume to be static, and one for the NFW halo (NFWhalo) 
that would have been there, had baryon compression not taken place.
The real neutrino overdensity should then lie somewhere 
between these two extremes.  

For the NFWhalo run, we use the parameters 
$M_{\rm vir} = 1 \times 10^{12} M_{\odot}$ 
and $c=12$.  These numbers are taken from the paper of Klypin, Zhao and 
Somerville (hereafter, KZS) \cite{bib:klypin2002},
from their ``favoured'' mass model of the Milky Way.  
Note that we are not using the $c$-$M_{\rm vir}$ relation (\ref{eq:cmvir}),
which, as we discussed before, is only a statistical trend.  
However,
the concentration parameter $c$ should still carry a redshift 
dependence {\it \'{a} la} equation (\ref{eq:cz}), in order to reproduce 
the correct time dependence of the density profile.

For MWnow, we adopt 
the total mass distribution (halo+disk+bulge) presented in Figure 3 of KZS, 
and fit it  approximately to a power law from $r=0$ to $20 \ {\rm kpc}$,
\begin{equation}
M_{\rm fit}(r,z=0) = 2 \times 10^{11} \left( \frac{r}{20 \ {\rm kpc}} 
\right)^{1.19367} 
\ {\rm M_{\odot}},
\end{equation}
where $M(r)$ means the total mass contained within a radius $r$.
We assume this fit to hold for
the region inside a {\it physical} radius of $20 \ {\rm kpc}$ at all times,
i.e., $M_{\rm fit}(r,z) = M_{\rm fit}(a r,z=0)$.  The region outside
of this $20 \ {\rm kpc}$ sphere is not  affected by baryonic
compression according to the KZS mass model 
(cf. their Figure 7) so we adopt the original NFW density 
profile outwardly from $20 \ {\rm kpc}$.  Thus, schematically, we have
\begin{eqnarray}
M(r,r<r_0) &=& M_{\rm fit}(r), \nonumber \\
M(r,r \geq r_0) &=& M_{\rm NFW}(r) - M_{\rm NFW}(r_0)+M_{\rm fit}(r_0),
\end{eqnarray}
where $r_0 = 20 \ a^{-1} \ {\rm kpc}$, $M_{\rm NFW}(r)$ is the mass 
contained in an NFW halo at radius $r$
[or $M_{\rm halo}(r)$
in equation (\ref{eq:nfwmass})], and
$M_{\rm fit} (r_0)  \simeq 2 \times M_{\rm NFW} (r_0)$ for the parameters
used in this analysis.

\begin{figure}
\begin{center}
\epsfxsize=15.7cm
\epsfbox{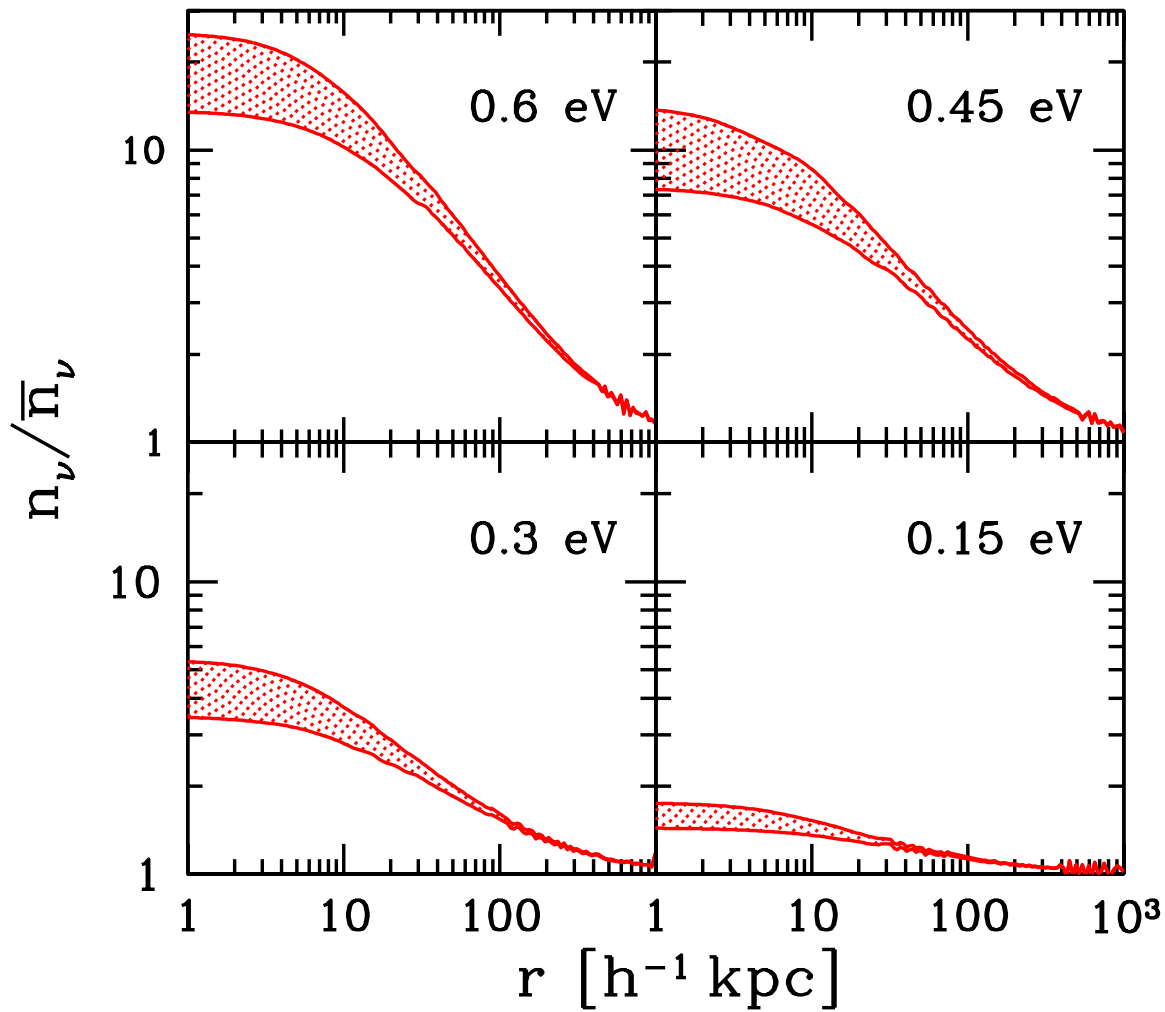}
\end{center}
\caption{\label{fig:milkyway}Relic neutrino number density per
flavour, $n_{\nu}=n_{\bar{\nu}}$, 
 in the Milky Way for various neutrino masses.  
All curves are normalised to 
$\bar{n}_{\nu} = \bar{n}_{\bar{\nu}} \simeq 56 \ {\rm cm}^{-3}$.  The
top curve in each plot corresponds 
to the MWnow run, and the bottom to the NFWhalo run.
The enclosed region represents a possible range of 
overdensities at $z=0$.}
\end{figure}

\subsection{Results and discussions}

Our Milky Way simulation results for four neutrino mass 
$m_{\nu}=0.15,0.3,0.45,0.6 \ {\rm eV}$
are displayed in Figure \ref{fig:milkyway}.  The 
shaded region in each plot corresponds to a possible range of  
overdensities at $z=0$. At first glance, it may seem unphysical that 
the apparently static MWnow potential (in {\it physical coordinates})
should capture so many neutrinos.  To resolve
this ``paradox'', one must remember that  neutrino clustering 
 is studied in the context of an expanding universe; the (unbound) 
neutrino thermal velocity decreases with time [equation (\ref{eq:fdspeed})], 
thus causing them to be more readily captured.  Equivalently,  
in  {\it comoving coordinates}, it is easy to see that while the neutrino 
conjugate momentum (\ref{eq:comoving}) does not redshift, the MWnow potential well
shrinks in size and deepens with time.

In each scenario we studied, 
the final momentum distribution at $r_{\oplus}$ is almost 
isotropic, with a zero mean radial velocity $\langle v_r \rangle$,
and second velocity 
moments that satisfy approximately the relation $2 \langle v_r^2 \rangle = 
\langle v_T^2 \rangle$ (cf. Table \ref{table:moments}).  For this reason,
we plot the smoothed, or coarse-grained, 
phase space densities $\bar{f}(r_{\oplus},p)$ only as 
functions of the absolute velocity (cf.
Figure \ref{fig:momentum}).

\begin{table}
\caption{\label{table:moments} Velocity moments at $r_{\oplus}$ for
various neutrino masses in the MWnow and NFWhalo runs 
(see text for definitions).  The first column shows the
overdensities $n_{\nu}/\bar{n}_{\nu}$.  The second, third and fourth
columns show the mean radial, transverse and absolute velocities in 
terms of the dimensionless quantities 
$\langle y_r \rangle$, $\langle y_T \rangle$  and 
$\langle y \rangle$, where $y = m_{\nu} v/T_{\nu,0}$.
In the last three columns are the second moments.
The corresponding values
for a relativistic Fermi--Dirac distribution are displayed in the first row.}
\begin{indented}
\item[]\begin{tabular}{@{}l|l|lll|lll}
\br  & $n_{\nu}/\bar{n}_{\nu}$ & $\langle y_r \rangle$ 
& $\langle y_T \rangle$ & 
$\langle y \rangle$ & $\langle y_r^2 \rangle$ &
$\langle y_T^2 \rangle$ & $\langle y^2 \rangle$
\\ \mr Relativistic\ Fermi--Dirac & 1&
$0$ & $2.48$ & $3.15$ & $4.31$ & $8.63$ & $12.94$ \\
\mr
NFWhalo, $m_{\nu}=0.6 \ {\rm eV}$ & 12 &
$0.0$ & $3.4$ & $4.3$ & $6.9$ & $13$ & $20$ \\
NFWhalo, $m_{\nu}=0.45 \ {\rm eV}$ & 6.4 &
$0.0$ & $2.8$ & $3.5$ & $4.6$ & $9.5$ & $14$ \\
NFWhalo, $m_{\nu}=0.3 \ {\rm eV}$ & 3.1 &
$0.0$ & $2.3$ & $3.0$ & $3.6$ & $7.3$ & $11$ \\
NFWhalo, $m_{\nu}=0.15 \ {\rm eV}$ & 1.4 &
$0.0$ & $2.3$ & $2.0$ & $3.8$ & $7.6$ & $11$ \\
\mr
MWnow, $m_{\nu}=0.6 \ {\rm eV}$ & 20 &
$0.0$ & $4.0$ & $5.1$ & $9.3$ & $18$ & $28$ \\
MWnow, $m_{\nu}=0.45 \ {\rm eV}$ & 10 &
$0.0$ & $3.1$ & $4.0$ & $6.1$ & $12$ & $18$ \\
MWnow, $m_{\nu}=0.3 \ {\rm eV}$ & 4.4 &
$0.0$ & $2.5$ & $3.2$ & $3.9$ & $8.0$ & $12$ \\
MWnow, $m_{\nu}=0.15 \ {\rm eV}$ & 1.6 &
$0.0$ & $2.3$ & $2.9$ & $3.7$ & $7.3$ & $11$ \\
\br
\end{tabular}
\end{indented}
\end{table}

\begin{figure}
\begin{center}
\epsfxsize=15.7cm
\epsfbox{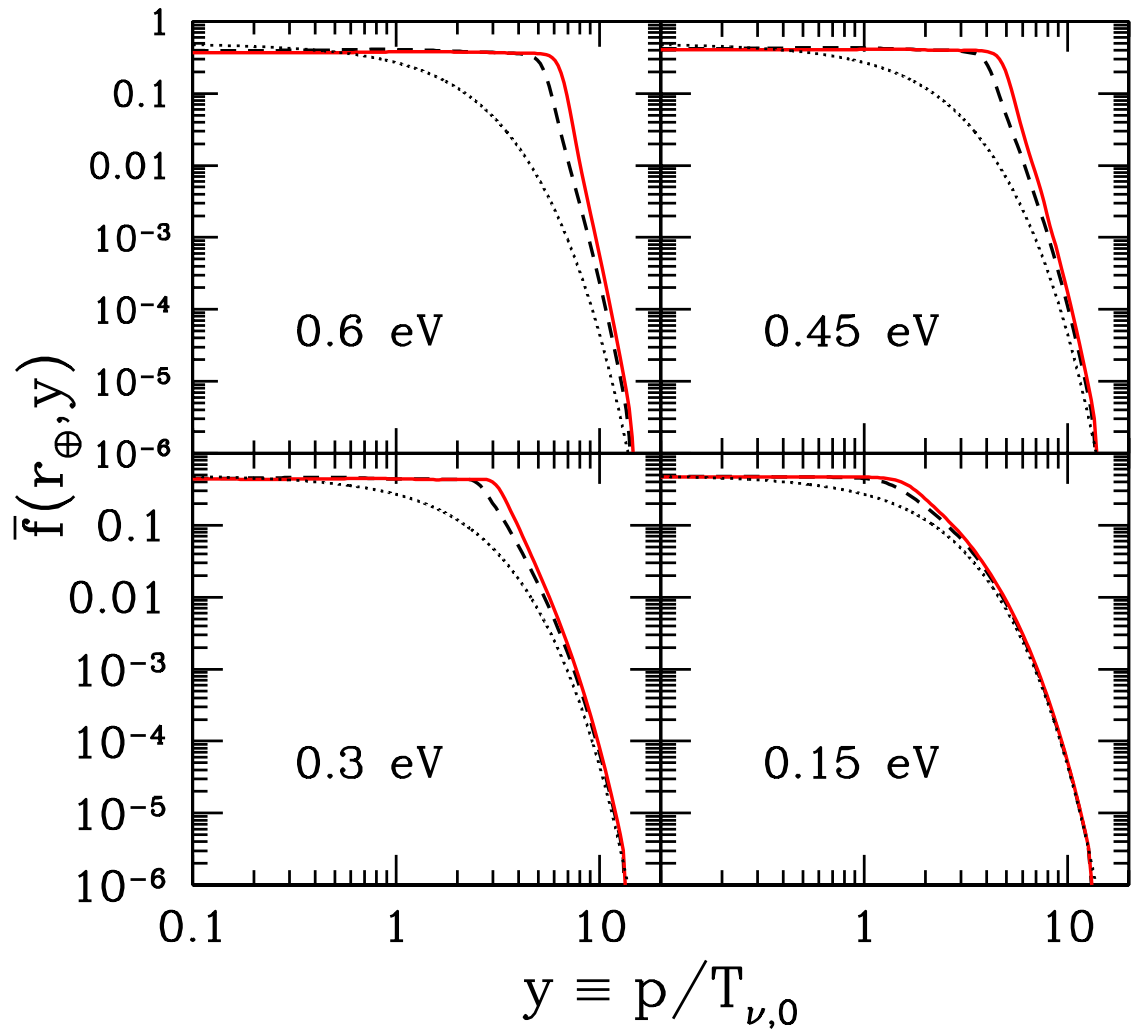}
\end{center}
\caption{\label{fig:momentum}Momentum distribution of relic neutrinos
at $r_{\oplus}$ for various neutrino masses.  
The red (solid) line denotes  the MWnow run,
while the dashed line represents the NFWhalo run.  The relativistic 
Fermi--Dirac function is indicated by the dotted line.  The 
escape velocity $v_{\rm esc} = \sqrt{2 |\phi(r_{\oplus})|}$ is 
$490 \ {\rm km \ s}^{-1}$ and $450 \ {\rm km \ s}^{-1}$ for MWnow and 
NFWhalo respectively, corresponding to ``escape momenta'' 
$y_{\rm esc} \equiv m_{\nu} v_{\rm esc}/T_{\nu,0}$
of $(5.9,4.4,3.0,1.5)$ and $(5.4,4.1,2.7,1.4)$ for 
$m_{\nu} = (0.6,0.45,0.3,0.15) \ {\rm eV}$.}
\end{figure}

As expected, the coarse-grained
distribution $\bar{f}(r_{\oplus},p)$ for the case with the highest overdensity
(MWnow, $m_{\nu}=0.6 \ {\rm eV}$) resembles the original Fermi--Dirac spectrum
the least, while $\bar{f}$ for the case with the lowest overdensity 
(NFWhalo, $m_{\nu}=0.15 \ {\rm eV}$) is almost Fermi--Dirac-like.  All
spectra share the feature that they are flat at low momenta, with a 
common value  of $\sim 1/2$.  The turning point for each distribution
coincides approximately with the ``escape momentum'' $p_{\rm esc}$  
(i.e., 
$m_{\nu}$ times the escape velocity 
$v_{\rm esc} = \sqrt{2 |\phi(r_{\oplus})|}$)  
for the system concerned, beyond which the phase space density falls off
rapidly, until it matches again the Fermi--Dirac function at the very
high momentum end of the spectrum.    Deviation from the original 
Fermi--Dirac spectrum is therefore most severe around $p_{\rm esc}$.  

The maximum value of $\bar{f}$ is a little less than $ 1/2$.  This
is consistent with the requirement that the final coarse-grained density
must not exceed the maximal value of the initial fine-grained 
distribution,  $\bar{f} \leq \max(f_0)$ \cite{bib:lyndenbell}.  
For neutrinos, $f_0$ has a value
of $1/2$ at $p=0$.  Thus, our $\bar{f}$ not only satisfies but completely
saturates the bound at low momenta up to $p_{\rm esc}$, forming
a kind of semi-degenerate state that can only be made denser by filling
in states above $p_{\rm esc}$.\footnote{This degeneracy 
should not be confused with that arising from the Pauli exclusion
principle.}  However, since neutrinos with momenta above $p_{\rm esc}$ do not 
become gravitationally bound to the galaxy/halo, these high momentum states
are much less likely to be fully occupied.  
This explains $\bar{f}$'s rapid drop beyond $p_{\rm esc}$.  Also,
the hottest neutrinos are not significantly affected by the galaxy/halo's
gravitational forces. Therefore the very high end of the momentum
distribution remains more or less Fermi--Dirac-like.  Finally, we note
that because the filling of phase space happens from bottom up,
the mean momenta for the least clustered cases tend to be lower
than the Fermi--Dirac value $\langle p \rangle \simeq 3.15 T_{\nu,0}$,
in contrast with the na\"{\i}ve expectation that clustering is necessarily
accompanied by an increase in the neutrinos' average kinetic energy.

\subsection{Tremaine--Gunn bound}

It is interesting to compare our results with nominal bounds
from phase space arguments.  By demanding the final coarse-grained
distribution to be always less than the maximum of the original 
fine-grained distribution, Tremaine and Gunn \cite{bib:tremaine&gunn}
argued in 1979 that if neutrinos alone are to 
constitute the dark matter of a galactic halo, their mass must be larger 
than $20 \ {\rm eV}$, assuming that the halo has a Maxwellian phase 
space distribution
as motivated by the theory of violent relaxation
\cite{bib:lyndenbell,bib:shu1978,bib:shu1987}.   A modern version of this
bound, in which the assumption about the phase space distribution is relaxed
and which allows for contribution to the total gravitational potential
from more than one form of matter, can be found in reference
 \cite{bib:kull}.
The revised mass bound may be written, alternatively, in the form of 
a constraint on the overdensity, which reads
\begin{equation}
\label{eq:kullbound}
\frac{n_{\nu}}{\bar{n}_{\nu}} 
< \frac{m_{\nu}^3 v_{\rm esc}^3}{9 \zeta(3) T_{\nu,0}^3}.
\end{equation}
For the  neutrino masses $m_{\nu}=(0.6,0.45,0.3,0.15) \ {\rm eV}$, 
this expression evaluates 
to $(19,8.0,2.4,0.3)$ and $(15,6.4,1.8,0.25)$  for MWnow and
NFWhalo, respectively, at $r_{\oplus}$.  

At first sight, some of our numerical
results seem to have completely violated the bound (\ref{eq:kullbound}). 
But this 
cannot be, since we have seen explicitly that all of our final coarse-grained
distributions satisfy perfectly the constraint $\bar{f} \leq \max(f_0)$.
Furthermore, an upper bound of 0.25 on the overdensity for a $0.15 \ {\rm eV}$
neutrino is obviously nonsensical.  

The answer, as the astute reader would have figured, 
lies in accounting.
In the derivation of (\ref{eq:kullbound}), a semi-degenerate distribution
has been summed only up to the momentum state corresponding to
the escape velocity of the system.  Neutrinos with higher momenta
that could be hovering around in the vicinity have been 
completely ignored.  In contrast, in our calculations, 
it is of no concern to us whether or not the relic neutrinos actually form 
bound states with the galaxy/halo.  Therefore it is more appropriate for 
us to sum every neutrino
in sight, rather than imposing a cut-off at $v_{\rm esc}$.  However,
if we had imposed such a cut-off, one can easily see from Figure
\ref{fig:momentum} that our 
overdensities would have just saturated the bound (\ref{eq:kullbound}), 
so there is no conflict.  Nonetheless, this illustrates how nominal bounds 
such as (\ref{eq:kullbound}) must be used with care.

\begin{figure}
\begin{center}
\epsfxsize=15.7cm
\epsfbox{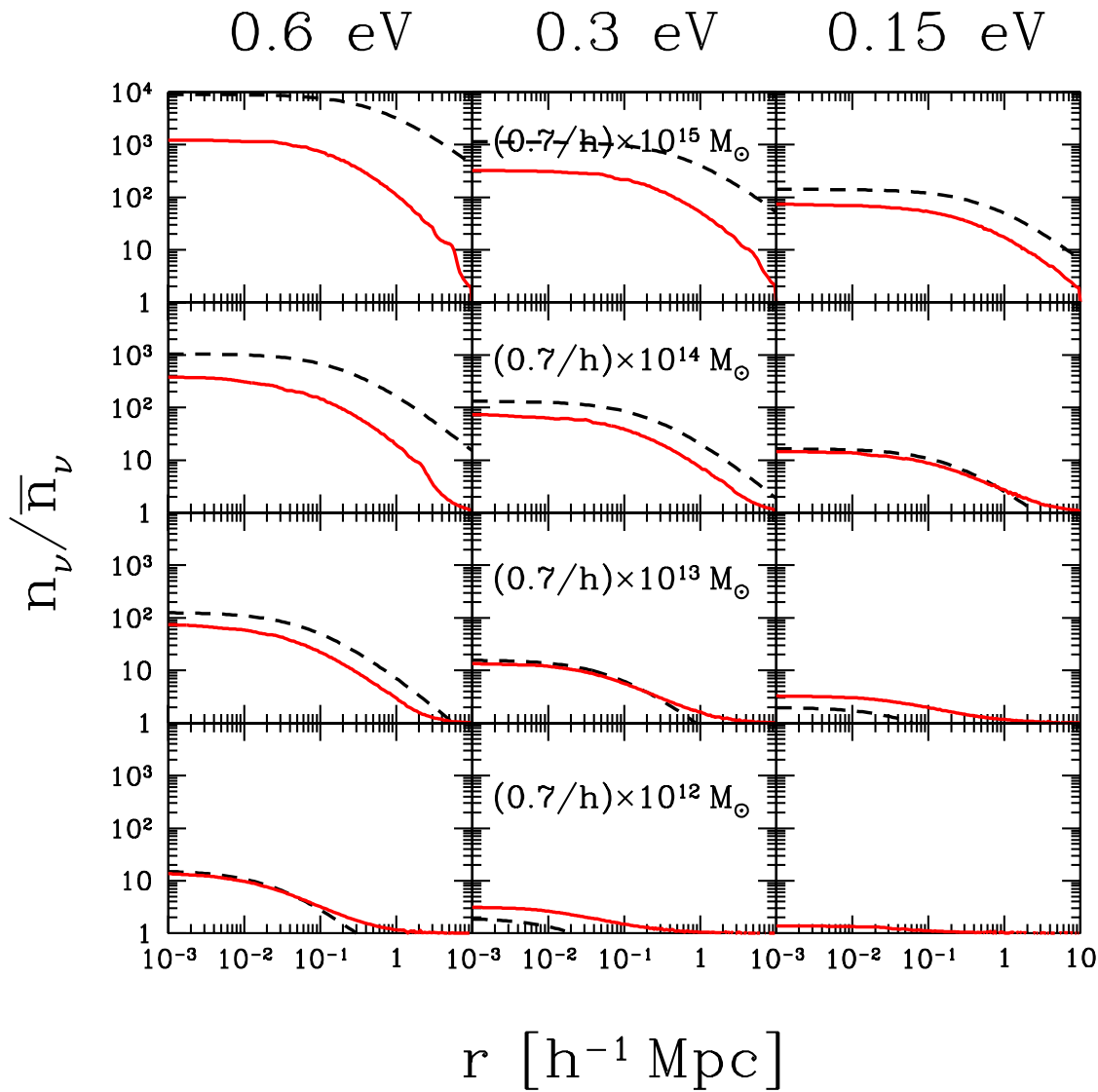}
\end{center}
\caption{\label{fig:tremainegunn}The Tremaine--Gunn bound on the neutrino
overdensity for various halo and neutrino masses (dashed lines).  The red
(solid) lines correspond to our $N$-$1$-body results from 
section~\ref{onebody}.}
\end{figure}

Before we conclude this section, let us note that for an NFW halo, the
bound (\ref{eq:kullbound}) can be written as
\begin{eqnarray}
\frac{n_{\nu}}{\bar{n}_{\nu}} & <& \frac{m_{\nu}^3}{9 \zeta(3) T_{\nu,0}^3}
\left[\frac{2 G M_{\rm vir}}{g(c)} \frac{\ln(1+r/r_s)}{r} \right]^{3/2}
\nonumber \\
&\simeq &0.58 \times \left(\frac{m_{\nu}}{\rm eV} \right)^3 
\left[ \left(\frac{M_{\rm vir}}{10^{12} h^{-1} M_{\odot}} \right)
\left(\frac{h^{-1} {\rm Mpc}}{r} \right) \frac{\ln(1+r/r_s)}{g(c)} 
\right]^{3/2},
\end{eqnarray}
where the function $g(c)$ is defined in equation (\ref{eq:gc}) in the 
Appendix.  Figure \ref{fig:tremainegunn} shows the bound as a function
of radius for the various halo and neutrino masses considered in
section~\ref{onebody}, along with the 
overdensities obtained from $N$-$1$-body simulations. 
We find the limit (\ref{eq:kullbound}) to be 
saturated for the lightest halo and neutrino masses (and also some apparent
violation of the bound due to different accounting).
This explains
why some of the overdensities exhibit an almost $m_{\nu}^3$ dependence
(cf. Figure \ref{fig:msquare}).  On the other hand,
the heaviest $\{m_{\nu},M_{\rm vir}\}$ set is short of the bound 
by at least one order of magnitude.
This is also consistent with expectations: heavier halo and neutrino masses 
give a higher escape momentum $p_{\rm esc}$, and
higher momentum states are more difficult
to fill up to a semi-degenerate level, since there are less particles in
these states in the original
Fermi--Dirac distribution to begin with.

\section{\label{detection} Implications for detection}
 
In this section, we determine the implications of our clustering 
results for the direct detection of relic neutrinos, 
in contrast with the purely cosmological inferences discussed
in section~\ref{cosm}. 
We consider various proposed detection methods based 
on scattering processes, involving the relic neutrinos 
both as a beam and as a target. In particular, we shall discuss 
(i) coherent elastic scattering of the relic neutrino flux 
off target matter in a terrestrial detector (section~\ref{flux}), 
as well as (ii) the scattering of extremely energetic particles 
(accelerator beams or cosmic rays) off the relic neutrinos as 
a target (section~\ref{target}).

\subsection{\label{flux} Flux detection} 
 
The low average momentum $\langle p\rangle =\langle y\rangle\, T_{\nu ,0}$
of the relic neutrinos 
(cf. Table~\ref{table:moments}) corresponds to a (reduced) 
de Broglie wavelength 
of macroscopic dimension,  
$\lambdabar = 1/\langle p\rangle = 0.12$~cm$/\langle y\rangle$   
(cf. Table~\ref{table:deBroglie}).  
Therefore, one may envisage scattering processes in which many target atoms
act  
coherently~\cite{Shvartsman:sn,Smith:jj} over a macroscopic volume 
$\lambdabar^3$, so that the reaction rate  
for elastic scattering becomes  
proportional to the square of the number of target atoms in that volume. 
Compared to the case where the neutrinos are 
elastically scattered coherently only on  
the individual nuclei of the target, 
the rate in this case is enhanced by a huge factor of
\begin{equation} 
\label{eq:enhancement} 
\frac{N_A}{A}\,\rho_{\rm t}\,\lambdabar^3 
\simeq 6\times 10^{18}\,\left( \frac{100}{A}\right)  
\left( \frac{\rho_{\rm t}}{{\rm g/cm^3}}\right) 
\left( \frac{\lambdabar}{0.1\ {\rm cm}}\right)^3  
\,, 
\end{equation} 
where $N_A$ is the Avogadro constant, $A$ is the atomic mass, and 
$\rho_{\rm t}$ is the 
mass density of the target 
material.\footnote{In the case of coherent scattering,  
it is possible, in principle, to measure also the scattering amplitude  
itself~\cite{Stodolsky:1974aq,Cabibbo:bb,Langacker:ih}, which 
is linear in the Fermi coupling constant $G_F$.  
However, one needs a large lepton asymmetry for 
a non-negligible  effect.} 
 
\begin{table} 
\caption{\label{table:deBroglie} Properties of the relic neutrinos at
$r_{\oplus}$ for
various neutrino masses from the MWnow and NFWhalo runs 
(see text for definitions)  relevant for their direct detection.  
The first column shows the
overdensities.  Columns two, three and four show, respectively,
 the mean absolute momenta, the associated mean reduced 
de Broglie 
wavelengths and the mean absolute velocities (in units of $c$).   
The corresponding values
for a relativistic Fermi--Dirac distribution are displayed in row one.} 
\begin{indented} 
\item[]\begin{tabular}{@{}l|l|l|l|l} 
\br & $\frac{n_\nu}{\bar{n}_\nu}$   & $\langle p \rangle$ & $\lambdabar = 
\frac{1}{\langle p \rangle}$   
 & $\langle v\rangle$  
\\ \mr Relativistic\ Fermi--Dirac & 1   & $5.3\times 10^{-4}$~eV & 
$3.7\times 10^{-2}$~cm  
  & see (\ref{eq:fdspeed})  \\ 
\mr 
NFWhalo, $m_\nu =0.6 \ {\rm eV}$ & 12 &  $7.2\times 10^{-4}$~eV &
$2.7\times 10^{-2}$~cm   & 
 $1.2\times 10^{-3}$  \\ 
NFWhalo, $m_\nu =0.45 \ {\rm eV}$  &  6.4 &    $5.9\times 10^{-4}$~eV & 
$3.4\times 10^{-2}$~cm  & 
 $1.3\times 10^{-3}$  \\ 
NFWhalo, $m_\nu =0.3 \ {\rm eV}$ &  3.1 &  $5.0\times 10^{-4}$~eV  & 
$3.9\times 10^{-2}$~cm   &
 $1.7\times 10^{-3}$  \\ 
NFWhalo, $m_\nu =0.15 \ {\rm eV}$ &  1.4 &  $3.4\times 10^{-4}$~eV  &  
$5.9\times 10^{-2}$~cm   & 
$2.2\times 10^{-3}$  \\ 
\mr 
MWnow, $m_\nu =0.6 \ {\rm eV}$ &  20 &  $8.5\times 10^{-4}$~eV  & 
$2.3\times 10^{-2}$~cm   & 
 $1.4\times 10^{-3}$  \\ 
MWnow, $m_\nu =0.45 \ {\rm eV}$  &  10 &  $6.7\times 10^{-4}$~eV  & 
$2.9\times 10^{-2}$~cm  & 
 $1.5\times 10^{-3}$  \\ 
MWnow, $m_\nu =0.3 \ {\rm eV}$  &  4.4 &   $5.4\times 10^{-4}$~eV & 
$3.7\times 10^{-2}$~cm  & 
 $1.8\times 10^{-3}$  \\ 
MWnow, $m_\nu =0.15 \ {\rm eV}$ &  1.6 &  $4.9\times 10^{-4}$~eV  & 
$4.1\times 10^{-2}$~cm   & 
 $3.2\times 10^{-3}$  \\ 
\br 
\end{tabular} 
\end{indented} 
\end{table}

By exploiting the above coherence effect, a practical
detection scheme for the local relic neutrino flux 
is based on the fact that a test body of density $\rho_{\rm t}$  
at Earth will experience a neutrino wind force through random neutrino 
scattering events, leading
to an acceleration~\cite{Shvartsman:sn,Smith:jj,Duda:2001hd,Ferreras:wf}
\begin{eqnarray} 
\label{eq:accel} 
a_{\rm t} &\simeq & \sum_{\nu,\bar\nu}\  
\underbrace{n_{\nu}\,v_{\rm rel}}_{\rm flux}\ 
\frac{4\pi}{3}\, N_A^2\, \rho_{\rm t}\,  r_{\rm t}^3 
\  
\sigma_{\nu N}\,  
\underbrace{2\,m_\nu\,v_{\rm rel}}_{\rm mom.\, transfer} 
\nonumber \\
&\simeq & 
2\times 10^{-28}\  
\left( \frac{n_\nu}{\bar n_\nu}\right)\,  
\left( \frac{10^{-3}\,c}{v_{\rm rel}}\right)\, 
\left( \frac{\rho_{\rm t}}{{\rm g/cm^3}}\right)\, 
\left( \frac{r_{\rm t}}{\hbar/(m_\nu v_{\rm rel})}\right)^3 
\ {\rm cm}\ {\rm s}^{-2}, 
\end{eqnarray}     
where  $\sigma_{\nu N}\simeq G_F^2\, m_\nu^2/\pi$ is the elastic  
neutrino--nucleon cross section, and
$v_{\rm rel}=\langle |\bm{v} - \bm{v}_\oplus|\rangle$ is the mean 
velocity of the relic neutrinos  
in the rest system of the  
detector. Here, $v_\oplus\simeq 2.3\times 10^2\ {\rm km \ s}^{-1} 
\simeq 7.7\times 10^{-4}\,c$ 
denotes  
the velocity of the Earth through the Milky Way.  Expression~(\ref{eq:accel})
is valid as long as the radius $r_{\rm t}$ of the target 
is smaller than the reduced de Broglie wavelength 
$\lambdabar = \hbar/(m_\nu v_{\rm rel})$ of the relic  
neutrinos. Furthermore, it applies only to
Dirac neutrinos. For Majorana neutrinos, the acceleration is suppressed,  
in comparison with~(\ref{eq:accel}), by a factor of 
$(v_{\rm rel}/c)^{2}\simeq 10^{-6}$  
for an unpolarised target, or $v_{\rm rel}/c\simeq 10^{-3}
$ for a polarised one.  
A target size much larger than $\lambdabar$ can be exploited, while  
avoiding destructive interference, by using 
foam-like~\cite{Shvartsman:sn}  or  
laminated~\cite{Smith:jj}  materials.  
Alternatively, grains of size $\sim\lambdabar$ could be randomly embedded 
(with spacing $\sim \lambdabar$) 
in a low density host material~\cite{Smith:1991mm,Smith:sy}.    
 
To digest these estimates, we note that the smallest 
measurable acceleration at present
is $\gwig 10^{-13} \ {\rm cm \ s}^{-2}$, using 
conventional Cavendish-type torsion balances. 
Possible improvements with currently available 
technology to a sensitivity of  
$\gwig 10^{-23} \ {\rm cm \ s}^{-2}$ have been 
proposed \cite{Hagmann:1998nz,Hagmann:1999kf}. However, such a  
sensitivity is still off the prediction~(\ref{eq:accel}) by at 
least three orders of magnitude,  
as an inspection of the currently allowed range of local relic neutrino 
overdensities  in Table~\ref{table:deBroglie} reveals.
Therefore, we conclude that an observation of this effect will not be 
possible  
within the upcoming decade, but can still be envisaged in the foreseeable 
future  
(thirty to forty years according to reference 
\cite{Smith:sy}, exploiting advances 
in nanotechnology), as long as 
our known light neutrinos are Dirac particles. Should they turn out, 
in the meantime, to be  
Majorana particles, flux detection via mechanical forces will be 
a real challenge.  

Let us note finally that the background contribution to 
 the acceleration~(\ref{eq:accel}) 
from  the solar $pp$ neutrinos [${\rm flux} \sim 10^{11} \ {\rm cm}^{-2} 
{\rm s}^{-1}$, $\langle E_{\nu} \rangle \sim 0.3 \ {\rm MeV}$ (e.g., \cite{bib:bahcall})], 
$a_t^{\nu\,{\rm sun}}\simeq 10^{-27}\ {\rm cm \ s}^{-2}$~\cite{Duda:2001hd}, 
may be rejected by directionality. 
The background from weakly interacting massive particles (WIMPs $\chi$, with 
mass $m_\chi$)~\cite{Duda:2001hd},
\begin{eqnarray}
\label{eq:accel_WIMP} 
a_{\rm t}^{\rm WIMP} &\simeq & 
\underbrace{n_{\chi}\,v_{\rm rel}}_{\rm flux}\ 
N_A\,A\
\sigma_{\chi N}\,  
\underbrace{2\,m_\chi\,v_{\rm rel}}_{\rm mom.\, transfer} 
\\ \nonumber 
&\simeq & 
6\times \! 10^{-29}
\left( \frac{\rho_\chi}{0.3\ {\rm GeV/cm^3}}\! \right) \!
\left( \frac{v_{\rm rel}}{10^{-3}\,c}\right)^2 \!
\left( \frac{A}{100}\right) \!
\left( \frac{\sigma_{\chi N}}{10^{-45}\ {\rm cm^2}}\! \right)
 {\rm cm}\ {\rm s}^{-2}, 
\end{eqnarray}
should
they be the main constituent of the galactic dark matter with mass density 
$\rho_\chi\equiv n_\chi m_\chi\simeq 0.3\ {\rm GeV \ cm}^{-3}$ at $r_\oplus$, 
can be neglected as soon as the WIMP--nucleon cross section $\sigma_{\chi N}$ 
is smaller than $\sim 3\times 10^{-45}$~cm$^2$. 
This should be well established by the time relic neutrino direct detection
becomes a reality. Note also that  neutrinos produced in the Earth's 
atmosphere do not contribute appreciably to $a_t$
because of their small flux; for
$E_{\nu} \sim 0.1 \to 10  \ {\rm GeV}$, the integrated flux is  
$\lwig 1 \ {\rm cm}^{-2} {\rm s}^{-1}$ (e.g., \cite{bib:honda}).

\subsection{\label{target} Target detection} 

Let us consider next the idea to take advantage of the fact that, 
for center-of-mass (c.m.) energies below the $W$- and $Z$-resonances, 
the weak interaction cross sections grow rapidly with energy. 
One may then contemplate the 
possibility to 
exploit a flux of extremely energetic particles---either from accelerator 
beams or from cosmic rays---for scattering on the relic neutrinos.
 
\begin{table} 
\caption{\label{table:accelerators} Beam parameters of forthcoming 
accelerators and expected interaction
rates with relic neutrinos.} 
\begin{indented} 
\item[]\begin{tabular}{@{}l|c|r|r|r|r|r||r} 
\br accelerator & $N$   & $A$ & $Z$ &$E_N$ [TeV]  & $L$ [km] & $I$ [A] & 
$\frac{R_{\nu A}}{\left[\frac{n_\nu}
{\bar n_\nu}\,\frac{m_\nu}{\rm eV}\right]}$ \\ 
\mr 
 & $p$ & 1 & 1 & $7$  &  $26.7$ &  $5.8\times 10^{-1}$   & 
$2\times 10^{-8}$~yr$^{-1}$\\
LHC &     &           &            &                           &       
&         &              \\
 & Pb & 208 & 82 & $574$  & $26.7$ &   $6.1\times 10^{-3}$      &    
$1\times 10^{-5}$~yr$^{-1}$  \\
\mr
 & $p$ & 1 & 1 &  $87.5$  & $233$ & $5.7\times 10^{-2}$  & 
$2\times 10^{-7}$~yr$^{-1}$\\
VLHC & &  &  &    &  &   & \\ 
 & Pb & 208 & 82 &  $7280$  & $233$ & $5.7\times 10^{-4}$  & 
$1\times 10^{-4}$~yr$^{-1}$\\ 
\mr
ULHC & $p$ & 1 & 1 &  $10^7$  & $4\times 10^4$ & $1.0\times 10^{-1}$  
& $10$~yr$^{-1}$\\ 
\br 
\end{tabular} 
\end{indented} 
\end{table} 

\subsubsection{\label{accelerator} At accelerators}

We start with a discussion of the prospects to detect interactions with 
the relic
neutrinos at forthcoming 
accelerator beams such as the 
LHC~\cite{LHC} and the VLHC~\cite{VLHC}, with 
beam energies $E_N$ ranging from $7 \ {\rm TeV}$ for the LHC running with 
protons, up 
to several thousands of TeV for the heavy ion option at the VLHC  
(cf. Table~\ref{table:accelerators}). 
At these beams, the attainable momentum transfers and c.m. energies,  
\begin{equation}
\sqrt{s}=  \sqrt{2\,m_\nu\,E_N}\simeq 4.5 \  
\left( \frac{m_\nu}{\rm eV}\right)^{1/2}\,
\left( \frac{E_N}{\rm 10\ TeV} \right)^{1/2} \ {\rm MeV}, 
\end{equation}
are so small that the cross sections for their interactions with the
relic neutrinos are  
enhanced by a factor $\sim A^2$ due to coherent elastic scattering over
 the size of the nucleus~\cite{Freedman:1973yd},   
and grow linearly with the beam energy,  
\begin{equation}
\label{eq:sigma_nuA}
       {\sigma_{\nu\, {}^A_Z N}} \simeq A^2\,G_F^2\,s\,/\pi \simeq  
       {3.4\times 10^{-43}\ A^2\,
\left( \frac{m_{\nu}}{\rm eV}\right)\,
\left( \frac{E_N}{\rm 10\ TeV}\right)
\ {\rm cm}^2}.  
\end{equation}
This leads to a scattering 
rate~\cite{Muller:1987qm,Melissinos:1999ew,Weiler:2001wc}
\begin{eqnarray}
\label{eq:acc_rate}
       { R_{\nu\, {}^A_Z N}} &\simeq & 
\sum_{\nu,\bar\nu}\ n_\nu\,\sigma_{\nu\, {}^A_Z N}\,L\ I/(Z\,e)
 \\[1.5ex]  &\simeq & 
       {2\times 10^{-8}
\left( \frac{n_\nu}{\bar n_\nu}\right)
       \left( \frac{m_{\nu}}{{\rm eV}}\right)
 \frac{A^2}{Z}\,
       \left( \frac{E_N}{{\rm 10\ TeV}}\right)\,
 \left( \frac{L}{100\ {\rm km}}\right)\,
       \left( \frac{I}{0.1\ {\rm A}}\right)\,
\ {\rm yr}^{-1}}, \nonumber
\end{eqnarray}
for a beam of particles ${}^A_Z N$, with charge $Z\,e$, length $L$  
and current $I$.  
In view of the currently allowed range of local relic neutrino overdensities  
displayed in Table~\ref{table:deBroglie}, and the beam parameters of the 
next generation of accelerators summarised in Table~\ref{table:accelerators}, 
the expected rate~(\ref{eq:acc_rate})  
is clearly too small to give rise to an observable effect in the 
foreseeable future. 
The extremely energetic $574 \ {\rm TeV}$ lead beam at the LHC will have 
less than $10^{-4}$ 
interactions per year with relic neutrinos, for $m_\nu \lwig 0.6 \ {\rm eV}$ 
and, correspondingly, $n_\nu/\bar n_ \nu \lwig 20$. 
Even a lead acceleration option
for the VLHC, with $E_N\simeq 7280 \ {\rm TeV}$ (or,
equivalently, $35 \ {\rm TeV}$ per nucleon) 
and a current $I\simeq 5.7\times 10^{-4} \ {\rm A}$ 
(a hundredth of the nominal 
current of the $p$ running mode)
will give  less than $10^{-3}$ events per year for the most optimistic 
neutrino mass scenario. 
Thus, there is little hope, in the foreseeable future, to detect relic
neutrino using terrestrial
accelerator beams. 

Let us nevertheless dream about the far future, in which an 
Ultimate Large Hadron Collider (ULHC) exists and is able to
accelerate protons to energies above $10^7 \ {\rm TeV}$\footnote{
Note that a collider at this energy has to be built anyhow if one wishes
to explore the ``intermediate'' scale 
$(M_{\rm EW}\,M_{\rm GUT})^{1/2}\sim 10^{10} \ {\rm GeV}$ between
the electroweak scale $M_{\rm EW}\sim 1 \ {\rm TeV}$ 
and the scale of grand unification $M_{\rm GUT}\sim 10^{17} \ {\rm GeV}$. 
The intermediate scale is exploited in many schemes of 
supersymmetry breaking (e.g., \cite{Ibanez:1998rf}) 
and in seesaw mechanisms for neutrino masses \cite{bib:seesaw,bib:seesaw2}.}  
in a ring of ultimate circumference  
$L\simeq 4\times 10^4 \ {\rm km}$ around the Earth, thus
leading to an interaction rate of more than one event per year 
(cf. Table~\ref{table:accelerators}).  
Even under these most optimistic circumstances, is it possible to 
reliably detect these interactions?  
Clearly, elastic scattering of the beam particles with the 
relic neutrinos---one of the contributions to the 
rate~(\ref{eq:acc_rate})---will be next to impossible to detect 
because of the small momentum transfers 
involved ($\sim 1 \ {\rm GeV}$ at $E_N\sim 10^7 \ {\rm TeV}$).  
A very promising alternative is to consider again a heavy ion beam, and 
to exploit the contribution of the inverse beta decay reaction,
\begin{equation}
\label{eq:inversebeta}
{}^A_Z N + \nu_e \to {}^A_{Z+1} N + e^- 
\,,
\end{equation}
to the rate~(\ref{eq:acc_rate}).
This reaction
changes the charge of the nucleus, causing it to follow 
an extraordinary trajectory and finally to exit the machine
such that it becomes susceptible
to detection~\cite{Melissinos:1999ew,Zavattini:unpubl}. 
A detection of this reaction would also clearly
demonstrate that a neutrino was involved in the scattering.

\subsubsection{\label{uhecr} With cosmic rays} 
 
In the meantime, until the ULHC has been constructed, target detection of
the relic neutrinos has to rely on extremely energetic cosmic rays. 
In fact, cosmic rays with up energies up to 
$E_{\rm cr}\sim 10^{20} \ {\rm eV}$
have been seen by air shower observatories.  The 
corresponding c.m. energies are
\begin{equation}
\sqrt{s}=  \sqrt{2\,m_\nu\,E_{\rm cr}}\simeq 14 \  
\left( \frac{m_\nu}{\rm eV}\right)^{1/2}\,
\left( \frac{E_{\rm cr}}{\rm 10^{20}\ eV} \right)^{1/2} \ {\rm GeV}
\end{equation}
when scattering off the relic neutrinos.  These energies 
are not too far  from the $W$- and $Z$-resonances, at which the 
electroweak cross sections become sizeable. Indeed, it was pointed out
long ago by 
Weiler~\cite{Weiler:1982qy,Weiler:1983xx} (for earlier suggestions, see 
\cite{Bernstein:1963,Konstantinov:1964,Cowsik:1964,Hara:1980,Hara:1980mz})
that the resonant annihilation of extremely energetic cosmic neutrinos 
(EEC$\nu$) 
with relic anti-neutrinos (and vice versa) into $Z$-bosons appears 
to be a unique process having sensitivity to the relic neutrinos. 
On resonance, 
\begin{equation}
E^{\rm res}_\nu=\frac{m_Z^2}{2m_\nu }\simeq 4\times 10^{21}\,
\left( \frac{\rm eV}{m_\nu }\right)
 \ {\rm eV},
\end{equation} 
the associated cross section is enhanced by several orders of magnitude, 
\begin{equation}
\langle\sigma_{\rm ann}\rangle =\int ds/m_Z^2\ \sigma_{\nu\bar\nu}^Z(s)\simeq 
2\pi\sqrt{2}\,G_F\simeq 4\times 10^{-32}\ {\rm cm}^2
,
\end{equation} 
leading to a ``short'' mean free path $\ell_{\nu}=(\bar n_{\nu}\,
\langle \sigma_{\rm ann}
\rangle)^{-1}\simeq 1.4\times 10^5 \ {\rm Mpc}$ which is {\it only} 
about $48\,h$ times the Hubble distance.
Neglecting cosmic evolution effects,
this corresponds to an annihilation probability for EEC$\nu$  
from cosmological distances 
on the relic neutrinos of $2\,h^{-1}\%$.

The signatures of annihilation are (i) absorption 
dips~\cite{Weiler:1982qy,Weiler:1983xx,bib:absorption} 
(see also \cite{Gondolo:1991rn,Roulet:1993pz,Yoshida:1997ie})
in the EEC$\nu$  spectrum at the resonant energies, and (ii) emission 
features~\cite{Fargion:1999ft,Weiler:1999sh,Yoshida:1998it,Fodor:2001qy,Fodor:2002hy} ($Z$-bursts) 
as protons (or photons) with energies spanning a decade or more above 
the predicted 
Greisen--Zatsepin--Kuzmin (GZK) cutoff at 
$E_{\rm GZK}\simeq 4\times 10^{19}$~eV~\cite{Greisen:1966jv,Zatsepin:1966jv}. 
This is the energy beyond which the CMB is absorbing to nucleons due to 
resonant photopion production.\footnote{The association of $Z$-bursts with 
the mysterious
cosmic rays observed above $E_{\rm GZK}$ is a controversial 
possibility~\cite{Fargion:1999ft,Weiler:1999sh,Yoshida:1998it,Fodor:2001qy,Fodor:2002hy,
Kalashev:2001sh,Gorbunov:2002nb,Semikoz:2003wv,Gelmini:2004zb}.}

The possibility to confirm the existence of relic neutrinos within 
the next decade from a measurement 
of the aforementioned absorption dips in the EEC$\nu$ 
flux was recently investigated in \cite{bib:absorption}. 
Presently planned neutrino detectors (Pierre Auger Observatory~\cite{Auger}, 
IceCube~\cite{IceCube}, ANITA~\cite{ANITA}, EUSO~\cite{EUSO}, 
OWL~\cite{OWL}, and SalSA~\cite{Gorham:2001wr})
operating in the energy regime above $10^{21} \ {\rm eV}$ appear to 
be sensitive enough 
to lead us, within the next decade, into an era of relic neutrino absorption 
spectroscopy, provided that the flux of the EEC$\nu$ 
at the resonant energies is close to current observational
bounds and the neutrino mass
is sufficiently large, 
$m_\nu\gwig\, 0.1 \ {\rm eV}$. In this case, the associated $Z$-bursts 
must also be seen as post-GZK events at the planned cosmic ray detectors
(Pierre Auger Observatory, EUSO, and OWL).

\begin{figure}
\begin{center}
\epsfxsize=15.7cm
\epsfbox{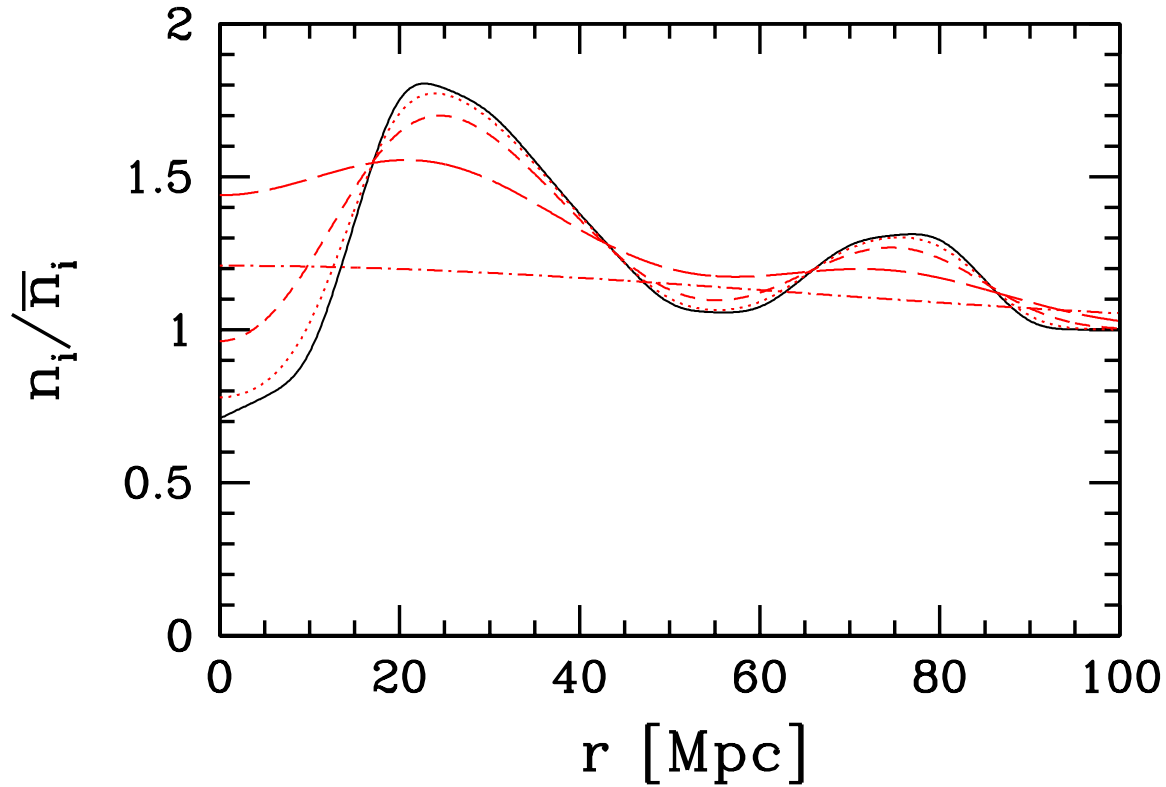}
\end{center}
\caption{\label{fig:local}``Large scale'' overdensities ($i=\nu,{\rm CDM}$) in the 
local universe, with the Milky Way at the origin,  estimated from 
equation (\ref{eq:nufilter}) assuming the function $K(\bm{x})$ to be 
a Gaussian.  The black (solid) line corresponds to the local CDM distribution
inferred from peculiar velocity measurements 
\cite{daCosta:1996nt} (see also \cite{Dekel:1998we}) 
smeared over the surface of a sphere with radius $r$ \cite{Fodor:2002hy}.  
The dotted line is the neutrino overdensity for
$m_{\nu} = 0.6 \ {\rm eV}$, short dash $0.3 \ {\rm  eV}$, long dash 
$0.15 \ {\rm eV}$, and dot-dash $0.04 \ {\rm eV}$.}
\end{figure}

What are the implications of relic neutrino clustering 
for absorption and emission spectroscopy? 
Firstly, absorption spectroscopy is predominantly sensitive to the relic
neutrino background at early times, with
the depths of the absorption dips determined largely by the higher number
densities at large redshifts ($z\gg 1$). 
Since neutrinos do not cluster significantly 
until after $z\lwig 2$, clustering at recent times can only show up as
secondary dips with such minimal widths in energy~\cite{Reiter:unpubl}  
that they do not seem likely to be resolved by planned observatories.

On the other hand, emission spectroscopy is directly sensitive to the 
relic neutrino content 
of the local universe ($z\lwig 0.01 \Leftrightarrow 
r_{\rm GZK} \lwig 50 \ {\rm Mpc}$).
However, since the neutrino density contrasts approximately track those of 
the underlying CDM above the neutrino 
free-streaming scale $k_{\rm fs}^{-1}$ 
(cf. section~\ref{linear}), 
it is clear that there cannot be a substantial neutrino overdensity over the 
whole GZK volume ($\sim r_{\rm GZK}^3$).  Indeed, if we take the 
linear fitting formula (\ref{eq:nufilter}), and apply it to the local CDM
distribution inferred from peculiar velocity measurements 
(with a $\sim 5 \ {\rm Mpc}$ smoothing}), 
one can see in Figure~\ref{fig:local} that the estimated
neutrino overdensity is always  $\lwig 2$. 
Hence the overall emission
rate cannot be significantly enhanced by gravitational clustering.  

Or we could imagine doing ``relic neutrino tomography'' of the
local universe, by exploiting the fact that there are several
galaxy clusters ($\! \gwig  10^{14} M_{\odot}$), such as
Virgo (distance $\sim 15 \ {\rm Mpc}$) and Centaurus ($\sim 45 \ {\rm Mpc}$),
within the GZK zone with significant
neutrino clustering (cf. Figure~\ref{fig:overdensities}). 
One could conceivably search for
directional dependences in the emission events as a signature of 
EEC$\nu$ annihilating on relic anti-neutrinos 
(and vice versa). 
For example, AGASA has an angular resolution of $\sim 2^{\circ}$ 
\cite{bib:agasa}. This is already sufficient to resolve the internal 
structures of, say, the Virgo cluster
(distance $\sim 15 \ {\rm Mpc}$,
$M_{\rm vir} \sim 8 \times 10^{14} M_{\odot}$)
which spans some $10^{\circ}$ across the sky.
Using our $N$-$1$-body clustering results in Figure~\ref{fig:overdensities},
the average neutrino overdensity along the line of 
sight towards and up to Virgo is estimated to be 
$\sim 45$ and $\sim 5$ for $m_{\nu} = 0.6 \ {\rm eV}$ and
$0.15 \ {\rm eV}$ respectively, given an angular resolution of 
$\sim 2^{\circ}$.  The corresponding increases in the number of events
coming from the direction of the Virgo cluster relative to the unclustered
case, assuming an isotropic distribution of EEC$\nu$ sources,
 are given roughly by the same numbers, since protons originating
from $\sim 15 \ {\rm Mpc}$ away arrive at Earth approximately unattenuated.
The numbers improve to $\sim 55$ and $\sim 8$ respectively with a finer
$\sim 1^{\circ}$ angular resolution.

Note that our estimates here are generally a factor of few higher than 
the predictions of SM.  This is expected, because the linear method
adopted in their analysis cannot account for additional clustering
from nonlinear effects, as we demonstrated in section~\ref{linear}.

\section{\label{conclusion} Conclusion}

We have conducted a systematic and exhaustive study of the gravitational
clustering of big bang relic neutrinos onto existing CDM and 
baryonic structures 
within the flat $\Lambda$CDM model, with the aim of understanding their 
clustering properties on galactic and sub-galactic scales for the purpose
of designing possible scattering-based detection methods.
Our main computational tools are (i) a restricted, $N$-$1$-body method 
(section~\ref{onebody}), in which we neglect the gravitational interaction
between the neutrinos and treat them as test particles moving in an
external potential generated by the CDM/baryonic structures,
and (ii) a semi-analytical, linear technique 
(section~\ref{linear}), 
which requires additional assumptions about the neutrino phase space 
distribution.  In both cases, the CDM/baryonic gravitational
potentials are calculated from parametric halo density profiles from 
high resolution $N$-body studies  \cite{Navarro:1995iw,bib:nfw} 
(section~\ref{profiles}) and/or from
realistic mass distributions reconstructed from observational data
(e.g., \cite{bib:klypin2002,daCosta:1996nt}).

Using these two computational techniques, we track
the relic neutrinos' accretion onto CDM halos ranging from the galaxy to the 
galaxy cluster variety ($M_{\rm vir} \sim 10^{12} \to 10^{15} M_{\odot}$), 
and determine the neutrino number densities
on scales $\sim 1 \to 1000 \ {\rm kpc}$ for neutrino masses satisfying
current constraints (\ref{eq:massbound}) from CMB and LSS
(Figures~\ref{fig:overdensities} and \ref{fig:nuvscdm}).  Because we can
simulate only a finite set of halo and neutrino parameters, we provide also
additional plots illustrating the approximate dependences of the neutrino 
overdensities on the halo and neutrino masses 
(Figures~\ref{fig:halomass} and \ref{fig:msquare}).  These can be used
for interpolation between simulation results.
Furthermore, we find that the linear technique systematically 
underestimates the neutrino overdensities over the whole range of
halo and neutrino masses considered in this study
(Figure~\ref{fig:overdensities}).  
Reconciliation with $N$-$1$-body simulations can only be achieved if
we impose a smoothing scale of $\gwig 1 \ {\rm Mpc}$, or if
the overdensity is no more than three or four.  
We therefore conclude that the linear theory does not generally 
constitute a faithful approximation to the Vlasov equation
in the study of neutrino clustering on  galactic and  sub-galactic scales 
($\! \lwig 50 \ {\rm kpc}$).  However, it may still be useful for finding the 
minimum effects of neutrino clustering in other contexts not considered in this
work (e.g., the nonlinear matter power spectrum \cite{bib:kevtalk}).

Next we apply our $N$-$1$-body method to calculate the relic neutrino 
number density in the Milky Way (section~\ref{milkyway}), 
especially  their phase space distribution in our local 
neighbourhood at Earth $r_{\oplus}$, taking also into
account contributions to the total gravitational potential
from the galactic bulge and disk.  
We find a maximum overdensity of $\sim 20$ per neutrino flavour 
in our immediate vicinity, provided that the neutrino mass is
at its current upper limit of $0.6 \ {\rm eV}$ (Table~\ref{table:moments} and
Figure~\ref{fig:milkyway}).  
For neutrino masses less than
$0.15 \ {\rm eV}$, the expected overdensity from gravitational clustering
is less than two.
The associated coarse-grained momentum spectra show varying degrees of
deviation from the relativistic Fermi--Dirac function,
but share a common feature that they are semi-degenerate, with phase space
density $\bar{f} \sim 1/2$,  up to the 
momentum state corresponding to the escape velocity from the 
Milky Way at $r_{\oplus}$
(Figure~\ref{fig:momentum}).  This means that the neutrino number densities
we have calculated here for $r_{\oplus}$ are already the 
{\it highest possible}, given the neutrino mass, without violating 
phase space constraints (e.g., \cite{bib:lyndenbell,bib:kull}).
In order to attain even higher densities, one must now appeal to
non-standard theories (e.g., \cite{Stephenson:1996qj}).

In terms of scattering-based detection possibilities, this meager enhancement
in the neutrino number density in the Milky Way from gravitational clustering  
means that relic neutrinos are still far from being detected in fully 
earthbound laboratory experiments.  For flux detection methods 
based on coherent elastic scattering of relic neutrinos off 
target matter in a terrestrial detector (section~\ref{flux}), a
positive detection could be thirty to forty years away \cite{Smith:sy},
provided that light neutrinos are Dirac particles.  For light
Majorana neutrinos, 
another $\sim 10^3$ times more sensitivity would be required in the detector
for a positive signal.  
Target detection methods using accelerator beams (section~\ref{accelerator})
seem equally hopeless, unless the accelerator is the size of the Earth
and operates at an energy of $\sim 10^7 \ {\rm TeV}$ 
(Table~\ref{table:accelerators}).

Meanwhile, target detection using extremely energetic cosmic neutrinos 
(EEC$\nu$, $\! \gwig 10^{21} \ {\rm eV}$) remains the only viable means to 
confirm the existence of big bang relic neutrinos
within the next decade or so (section~\ref{uhecr}).  Resonant
annihilation of EEC$\nu$ on relic neutrinos can be revealed as
absorption dips in the EEC$\nu$ flux (e.g., \cite{bib:absorption}), or
as emission features in the $Z$-decay products.  However, since absorption 
spectroscopy is largely insensitive to late time ($z \lwig 2$) 
relic neutrino clustering,
our findings here have little impact on the conclusions of 
\cite{bib:absorption}.  On the other hand, emission spectroscopy is
sensitive to the relic neutrino content of the local GZK zone,
$V_{\rm GZK} \sim 50^3 \ {\rm Mpc}^3$.  
While we find no significant large scale clustering 
within $V_{\rm GZK}$ (Figure~\ref{fig:local}) and therefore no significant
enhancement in the overall emission rates, it is still conceivable to  
exploit the considerable neutrino overdensities in nearby galaxy clusters, 
and search for directional dependences in the  post-GZK emission events.
For the Virgo cluster, for example, we estimate the event rate from
the central $1^\circ$ region to be
$\sim 55$ and $\sim 8$ times the unclustered rate for neutrino mass
$m_{\nu} = 0.6 \ {\rm eV}$ and $0.15 \ {\rm eV}$ respectively, assuming
an isotropic distribution of EEC$\nu$ sources.
[Our estimates differ by a factor of a few
from the predictions of \cite{bib:singh&ma}
because their linear technique fails to account for additional
clustering from nonlinear effects, as we have demonstrated in this study 
(section~\ref{linear})]. Planned observatories such as the Pierre Auger
Observatory \cite{Auger},  EUSO \cite{EUSO} and OWL \cite{OWL} 
will have sufficient angular resolution to, in principle, see this
enhancement. However, considering the rapidly improving constraints
on both the EEC$\nu$ flux and neutrino masses, it remains to be 
seen if the enhancement can indeed be observed 
with enough statistical significance.

\ack

We would like to thank David Cerde\~{n}o, Pasquale Di Bari, Birgit Eberle, 
Jan Hamann, Oleg Lebedev, Helmut Mais, and Thomas Reiter for useful 
discussions.

\appendix

\section{Details of $N$-$1$-body simulations}

We give in this appendix a detailed discussion of the techniques used in
our $N$-$1$-body simulations.

\subsection{Hamilton's equations}

Following \cite{bib:bertschinger}, we begin with a Lagrangian
$L (\bm{x},\dot{\bm x},\tau) = a  \left[ \frac{1}{2} m_{\nu} v^2 - m_{\nu} \phi
(|\bm{x}|,\tau) \right]$
for a test particle (neutrino) with mass $m_{\nu}$ and peculiar velocity
$\bm{v} \equiv \dot{\bm x}$ moving in a spherically symmetric
gravitational potential well $\phi(\bm{x},\tau)$ generated
by the CDM halo. Because of spherical symmetry, the
motion of the test particle is confined to a plane.  This allows
us to switch to polar coordinates $\{r,\theta\}$ to obtain
\begin{equation}
L (r,\theta, \dot{r},\dot{\theta},\tau) = a  \left[ \frac{1}{2} m_{\nu}
( \dot{r}^2 + r^2 \dot{\theta}^2 ) - m_{\nu} \phi (r,\tau) \right],
\end{equation}
and eliminate two superfluous variables in the process. The
canonical momenta conjugate to $r$ and $\theta$ are 
$p_r  \equiv  \partial L/\partial \dot{r} = a m_{\nu} \dot{r}$ and
$\ell \equiv r p_{\theta}  \equiv  \partial
L/\partial \dot{\theta} = a m_{\nu} r^2 \dot{\theta}$
respectively, leading to the Hamiltonian
\begin{equation}
H (r,\theta, p_r,\ell,\tau) = \frac{1}{2 a m_{\nu}} \left[p_r^2 +
\frac{\ell^2}{r^2} \right] + a m_{\nu} \phi(r,\tau),
\end{equation}
and hence Hamilton's equations
\begin{eqnarray}
\label{eq:hamilton} \frac{d r}{d \tau}  = \frac{p_r}{a m_{\nu}}, &
\qquad & \frac{d \theta}{d \tau}  = \frac{\ell}{a m_{\nu} r^2},
\nonumber
\\ \frac{d p_r}{d \tau}  = \frac{\ell^2}{a m_{\nu} r^3} - a m_{\nu}
\frac{\partial \phi}{\partial r}, &\qquad &  \frac{d \ell}{d \tau}
= 0,
\end{eqnarray}
where the last equality expresses the conservation of conjugate
comoving angular momentum $\ell$.

For computational purposes, it is convenient to rewrite the set of
differential equations (\ref{eq:hamilton}) in terms of redshift
$z$ and the normalised quantities
$u_{r} \equiv p_r/m_{\nu} = a \dot{r}$ and $u_{\theta} \equiv
\ell/m_{\nu} = a r^2 \dot{\theta}$,
thus rendering (\ref{eq:hamilton}) into the form
\begin{eqnarray}
\label{eq:normalised} \frac{d r}{d z}  =  - \frac{u_r}{\dot{a}}, &
\qquad & \frac{d \theta}{d z}  = - \frac{u_{\theta}}{\dot{a} r^2},
\nonumber
\\ \frac{d u_r}{d z}  =  -\frac{1}{\dot{a}} \left(\frac{u_{\theta}^2}{r^3}
-a^2 \frac{\partial \phi}{\partial r} \right),& \qquad & \frac{d
u_{\theta}}{d z} = 0,
\end{eqnarray}
where $a = (1+z)^{-1}$ and $\dot{a} = H_0 \sqrt{a^{-1} \
\Omega_{m,0} + a^2 \ \Omega_{\Lambda,0}}$
are, respectively, the scale factor and the expansion rate, the latter of which
is determined by the present day Hubble constant $H_0$ and the
cosmological model in hand.

The gravitational potential $\phi(r,\tau)$ obeys the Poisson equation
\begin{equation}
\nabla^2 \phi \to \frac{1}{r^2} \frac{\partial}{\partial r}
\left(r^2 \frac{\partial \phi}{\partial r} \right)= 4 \pi G a^2
\bar{\rho}_m(\tau) \delta_m(r,\tau).
\end{equation}
Since we are treating the CDM halo as a density 
perturbation sitting on top of
a uniform background, i.e., 
$\bar{\rho}_m(\tau) \delta_m(r,\tau) \to \rho_{\rm halo}(r)$, we have the 
following (partial) solution:
\begin{equation}
\frac{\partial \phi}{\partial r} =\frac{4 \pi G a^2}{r^2} \int^r_0
\rho_{\rm halo} (r',\tau) r'^2 dr'= \frac{G}{a r^2} M_{\rm halo} (r),
\end{equation}
where $M_{\rm halo}(r)$ is the physical mass contained in a sphere
of radius $r$. For an NFW halo,
\begin{eqnarray}
M_{\rm halo}(r)  =  4 \pi a^3 \rho_s r_s^3 g(x) 
= M_{\rm vir} g(x)/g(c), \qquad x=r/r_s,\label{eq:nfwmass} \\
g(x) =  \ln(1+x)-\frac{x}{1+x},\label{eq:gc} \\
c  = \frac{9}{1+z} \left(\frac{M_{\rm vir}}
{1.5 \times 10^{13} h^{-1} M_{\odot}} 
\right)^{-0.13}, \\
r_s  = \frac{1}{c} \left( \frac{3}{4 \pi} \frac{M_{\rm vir}}{\bar{\rho}_{m,0} 
\Delta_{\rm vir}} \right)^{1/3} \label{eq:rs} \nonumber \\
\hspace{5mm}  = 9.51 \times 10^{-1} \ c^{-1} \Omega_{m,0}^{-1/3} 
\Delta_{\rm vir}^{-1/3}
\left(\frac{M_{\rm vir}}{10^{12} h^{-1} M_{\odot}} \right)^{1/3} \ 
h^{-1} {\rm Mpc}.
\end{eqnarray}
The meanings of these various symbols are explained in section~\ref{profiles}.

\subsection{Discretising the initial phase space distribution}

The number of neutrinos at some initial time $\tau_i$ in the interval 
$(\bm{x},\bm{p}) \to( \bm{x} + d \bm{x},\bm{p}+d \bm{p} )$ is defined
to be
\begin{equation}
dN =  f(\bm{x},\bm{p},\tau_i)  d^3x \ d^3p.
\end{equation}
We assume the initial phase space distribution to be isotropic and
homogeneous, i.e., $f(\bm{x},\bm{p},\tau_i) = f_0(p)$. This initial
homogeneity implies that the ensemble's subsequent evolution under
a spherically symmetric potential will depend only on the initial 
radial distance $r$, radial momentum $p_r$, and transverse momentum 
$p_T= \sqrt{p^2 - p_r^2}$. Thus we may rewrite the 
phase space volume element as
\begin{equation}
d^3x \to r^2 \sin \theta \ d \theta \ d \phi \ d r, 
\qquad  d^3 p \to  p_T \ d p_T \ d p_r \ d \varphi,
\end{equation} 
with $r \in [0,\infty)$,  
$\theta \in [0,\pi]$, $\phi \in [0,2 \pi)$, and 
$p_T \in [0,\infty)$, 
$p_r \in (-\infty,\infty)$,
$\varphi \in [0,2 \pi)$.

In an ideal and perfectly deterministic calculation, one would track 
the motion of every particle for every occupied combination of $r$, $p_r$ 
and $p_T$.  This is obviously impossible in practice; one must therefore 
resort to simulating only  a representative set of particles
and endow each 
with a ``weight'' according to the phase space region from which the particle
is drawn.  We show below how this
weight is calculated.\footnote{We note for completeness that the concept 
of super particles with equal 
masses or, in our language, weights is adopted in most full scale 
$N$-body simulations \cite{bib:hockney}.  
The representative 
point particles are drawn randomly from the initial phase space distribution.}

Suppose that a point particle initially at $(r,p_r,p_T)$ 
is representative for all particles in the initial
phase space interval $(r_a,p_{r,a},p_{T,a}) \to (r_b,p_{r,b},p_{T,b})$.  The
weight carried by this representative particle is defined as
\begin{equation}
w_i \equiv \int^{r_b,p_{r,b},p_{T,b}}_{r_a,p_{r,a},p_{T,a}}
 \int_{\theta,\phi,\varphi}  dN,
\end{equation}
where $\int_{\theta,\phi,\varphi}$ means summing over all $\theta$,
$\phi$ and $\varphi$, which is simple:
$\int_{\theta,\phi,\varphi}  \sin \theta \
d \theta \ d \phi \ d \varphi  = 8 \pi^2$.
The spatial integral over $r$ is also readily calculable.
The remaining phase space integral $\int^{p_{r,b},p_{T,b}}_{p_{r,a},p_{T,a}}
f(p) p_T d p_T d p_r$ can be solved by way of the parameterisation
$p_r=p \cos \psi$, and $p_T = p \sin \psi$, and hence
\begin{equation}
p_T \ d p_T d p_r \to p^2 \sin \psi \ d \psi, \qquad p \in [0,\infty), \;
\psi \in [0,\pi).
\end{equation}
Thus the weight carried by a point particle centred on and representing
the interval $(r_a,p_a,\psi_a) \to (r_b,p_b,\psi_b)$ is given by
\begin{equation}
w_i  =  8 \pi^2 T_{\nu,0}^3 
\int^{r_b}_{r_a} r^2 d r
\int^{y_b}_{y_a}  f(y) y^2 d y \int^{\psi_b}_{\psi_a} \sin \psi d \psi,
\end{equation}
where $y = p/T_{\nu,0}$ is a dimensionless momentum variable.

\subsection{Kernel method for density profile estimation}

We construct the neutrino density profiles from the discrete outputs
of our $N$-$1$-body simulations using the kernel method of 
\cite{bib:reed} and \cite{bib:merritt}, which we reproduce here for 
completeness.

The number density corresponding to a set of particles is estimated to be
\begin{equation}
n(\bm{r})= \sum^N_{i=1} \frac{w_i}{h^3} K \left[\frac{1}{h}
\left|\bm{r} - \bm{r}_i \right| \right],
\end{equation}
where $w_i$ is the weight carried by the $i$th particle, $h$ is the 
window width, and $K$ is a normalised kernel such as the
Gaussian kernel,
\begin{equation}
K(y) = \frac{1}{(2 \pi)^{3/2}} e^{-y^2/2}.
\end{equation}
In order to obtain a spherically symmetric profile, we smear every particle 
around the surface of a sphere with radius $r_i$ centred on $r=0$.  
The spherically symmetric density estimate is then
\begin{equation}
n(r) = \sum^N_{i=1} \frac{w_i}{h^3} \widetilde{K} (r,r_i,h),
\end{equation}
with
\begin{equation}
\widetilde{K}(r,r_i,h) \equiv  \frac{1}{4 \pi} \int d \phi \int d
\theta \ \sin \theta  \widetilde{K} \left(h^{-1} \sqrt{r^2 + r_i^2
- 2 r r_i \cos \theta} \right).
\end{equation}
Substituting for the Gaussian kernel, we get
\begin{equation}
\widetilde{K} (r,r_i,h) = \frac{1}{2 (2 \pi)^{3/2}} \frac{h^2}{r
r_i} \left[e^{-(r-r_i)^2/2 h^2} - e^{-(r+r_i)^2/2h^2} \right].
\end{equation}
The optimal choice of $h$ generally depends on the underlying function
$n(r)$ \cite{bib:merritt}.  In our analysis, however, a 
constant $h$ is adopted for simplicity.  Our momentum distributions
in section~\ref{milkyway} are also constructed with this kernel method.

\end{document}